\documentclass[11pt,a4paper,twocolumn]{article}
\usepackage[utf8]{inputenc}
\usepackage{times}
\usepackage[colorlinks=false, pdfborder={0 0 0}]{hyperref}
\usepackage{breakurl}
\usepackage[rightcaption]{sidecap}
\usepackage{amsmath}
\usepackage{amssymb}
\usepackage{listings}
\usepackage{afterpage}
\usepackage[multiple]{footmisc}

\usepackage{enumitem}

\usepackage[font=small,margin=1ex,labelfont=bf]{caption}

\usepackage{subfigure}
\usepackage{calc}
\usepackage[left=2.0cm,right=2.0cm,top=3.0cm,bottom=3.0cm]{geometry}

\allowdisplaybreaks

\usepackage{graphicx}

\usepackage{fancyhdr}
\fancypagestyle{firstpage}{
  \fancyhead{}
  \cfoot{}
  \chead{\footnotesize{\em Technical Report No.~2017-01, pp.~15-28, Hochschule Niederrhein, Fachbereich Elektrotechnik \& Informatik (2017)}}
  
}

\title{\vspace*{-10mm}Construction of Confidence Intervals}
\author{
Christoph Dalitz\\
Institute for Pattern Recognition\\
Niederrhein University of Applied Sciences\\
Reinarzstr. 49, 47805 Krefeld, Germany\\
{\tt christoph.dalitz{@}hsnr.de}
}
\date{}

\begin{document}

% a), b), c) fur subitems
%\renewcommand{\labelenumi}{\alph{enumi})}
\renewcommand{\labelenumi}{\arabic{enumi})}

\twocolumn[
  \begin{@twocolumnfalse}
    \maketitle

\begin{abstract}%\noindent
Introductory texts on statistics typically only cover the classical ``two sigma'' confidence interval for the mean value and do not describe methods to obtain confidence intervals for other estimators. The present technical report fills this gap by first defining different methods for the construction of confidence intervals, and then by their application to a binomial proportion, the mean value, and to arbitrary estimators. Beside the frequentist approach, the likelihood ratio and the highest posterior density approach are explained. Two methods to estimate the variance of general maximum likelihood estimators are described (Hessian, Jackknife), and for arbitrary estimators the bootstrap is suggested. For three examples, the different methods are evaluated by means of Monte Carlo simulations with respect to their coverage probability and interval length. R code is given for all methods, and the practitioner obtains a guideline which method should be used in which cases.
\end{abstract}
\vspace*{2ex}

  \end{@twocolumnfalse}
  ]

\thispagestyle{firstpage}

\section{Introduction}
%-----------------------------------------------------------------------
When an unknown model parameter is estimated from experimental data,
the estimation always yields a value, be the sample size large or small.
We would, however, expect a more accurate value from a larger sample.
A {\em confidence interval} measures this ``accuracy'' in some way.
As ``accuracy''
can be defined in different ways, there are different approaches to the
construction of confidence intervals.

The most common approach is the {\em frequentist approach}, which is
based on the {\em coverage probability} and is taught in introductory
texts on statistics \cite{fahrmeir04}. It assumes the unknown parameter
to be known and then chooses an interval around the estimator that includes
the parameter with a given probability (typically $95\%$). The
{\em evidence based approach} utilizes the {\em likelihood ratio} and
chooses an interval wherein the likelihood function is greater than
a given threshold (typically $1/8$ of its maximum value) \cite{blume02}.
The {\em Bayesian approach} treats the unknown parameter as a random
variable and estimates its distribution from the observation. This leads
to the {\em highest posterior density} interval \cite{turkkan93}.

Both for binomial proportions and for mean values, simple formulas or algorithms
to compute confidence intervals can be given. A possible evaluation criteria
for the obtained intervals is the coverage probability. One should think
that this criterion favors the frequentist approach, but even for
this approach, the coverage probability may vary considerably, depending
on the true parameter value. For non-symmetric intervals, another evaluation
criterion is the interval length because, from two intervals with the same
coverage probability, the shorter one is preferable.

Beyond the binomial proportion and the mean value, there is no standard
formula for computing a confidence interval. For maximum likelihood
estimators, it is however known that they are asymptotically normal,
provided the likelihood function is sufficiently smooth \cite{greene00}.
In these cases, the confidence interval for the mean value can be used.
This requires an estimation of the estimator variance, which can be done
in two ways: the diagonal elements of the inverted {\em Hessian matrix}
of the log-likelihood function, or the {\em Jackknife} variance.

For non-smooth likelihood functions or for arbitrary estimators, only
the {\em bootstrap} method is universally applicable. This method
generates new data from the observations by random sampling with
replacement and estimates the confidence interval from the sampled data.
In principle, the bootstrap method is always applicable, even in cases
when the other methods work, but in the experiments described in this
report, the bootstrap method had a poorer coverage probability than the
classic confidence interval, and it should therefore only be used when 
other methods cannot be applied.

This report is organized as follows: section
\ref{sec:grundbegriffe} defines the basic terms estimator,
coverage probability, likelihood ratio, and posterior density.
In sections \ref{sec:p} and \ref{sec:mu}, the different approaches
are applied to the binomial proportion and to the mean value.
Sections \ref{sec:ml} and \ref{sec:bootstrap} describe construction
methods for confidence intervals for maximum likelihood estimators
and for arbitrary estimators.
Section \ref{sec:vergleich} presents Monte Carlo experiments that
evaluate the coverage probability of the different confidence intervals.
The final section makes recommendations which confidence interval should
be used in which case.

\section{Basic terms}
%-----------------------------------------------------------------------
\label{sec:grundbegriffe}
The probability distribution of a random variable $X$ be known except for
the value of some parameter $\theta$. In other words: the shape of
the probability density $f_\theta(x)$ be known, but not the value of
the parameters $\theta$. In the most general case, $\theta$ is a vector and
represents several parameter values. If $X$ is normal distributed, for
instance, then $\theta$ represents two parameters: $\theta=(\mu,\sigma^2)$.
An {\em estimator} is a function to estimate the unknown parameter from
independent observations $x_1,\ldots,x_n$ of the random variable $X$.
The particular estimated value is denoted with $\hat{\theta}$:
\begin{equation}
\label{eq:theta}
\hat{\theta} = \hat{\theta}(x_1,\ldots,x_n)
\end{equation}
Simple examples are the relative frequency as an estimator for a binomial
proportion, or the statistical average as an estimator for the parameter $\mu$
of the normal distribution.

\subsection{Maximum likelihood (ML)}
%-----------------------------------------------------------------------
\label{sec:grundbegriffe:ml}
The {\em maximum likelihood principle} is a general method to obtain
estimators \cite{greene00}. It chooses the parameter $\theta$ in such a
way that the {\em likelihood function} $L$ or\footnote{Note that $L(\theta)$ and $\log L(\theta)$ have their maximum at the same argument, because the logarithm is a monotonic function.}
the {\em log-likelihood function} $\ell$ is maximized:
\begin{subequations}
\label{eq:L}
\begin{eqnarray}
\label{eq:L:a}
L(\theta) & = & \prod_{i=1}^n f_\theta(x_i) \\
\label{eq:L:b}
\ell(\theta) & = & \log L(\theta) = \sum_{i=1}^n \log f_\theta(x_i)
\end{eqnarray}
\end{subequations}
Loosely speaking, $L(\theta)$ is a measure for the probability of the
observation $x_1,\ldots,x_n$ under the assumption that the true parameter
value is $\theta$. If $\theta=(\theta_1,\ldots,\theta_t)$ and
$\ell(\theta)$ is differentiable, the maximum likelihood principle yields
$t$ equations for the determination of the $t$ parameters
$\theta_1,\ldots,\theta_t$:
\begin{equation}
\label{eq:ML}
\frac{\partial}{\partial\theta_i} \ell(\theta) = 0 \quad\mbox{for }i=1\ldots,t
\end{equation}
Maximum likelihood estimators have a number of attractive properties like
asymptotic normality under quite general conditions. This will play a role
in section \ref{sec:ml}. In many cases, the equations (\ref{eq:ML}) cannot
be solved in closed form, thereby making a numerical maximization
of the log-likelihood function necessary. If this is not possible, one might
try other methods that possibly yield estimators in a simpler way, like
the {\em method of moments} or its generalization \cite{zsohar10}.

\subsection{Coverage probability}
%-----------------------------------------------------------------------
\label{sec:grundbegriffe:pcov}
An estimation function (\ref{eq:theta}) yields only a single value
and is therefore called a {\em point estimator}.
A {\em confidence interval}, on the contrary, gives a region
$[\theta_l, \theta_u]$ wherein the parameter falls with high probability.
The boundaries $\theta_{l,u}$ of the interval depend on the observed data
$x_1,\ldots,x_n$ and are thus random variables. The {\em frequentist approach}
is based on the following consideration: if $\theta$ is the true parameter
value, then it ideally should fall into the confidence interval with
a pre-defined  {\em coverage probability} $(1-\alpha)$:
\begin{equation}
\label{eq:Pcov}
P_{cov}(\theta) = P(\theta\in[\theta_l,\theta_u]) = 1-\alpha
\end{equation}
Unfortunately, Eq.~(\ref{eq:Pcov}) cannot be used to determine $\theta_l$ and
$\theta_u$, because the unknown $\theta$ is part of the equation. This dilemma
can be resolved when the problem is re-interpreted as a hypothesis testing
problem: under the hypothesis $\theta\notin[\theta_l,\theta_u]$,
the probability that the estimator deviates from $\theta$ more than the
observed value $\hat{\theta}$ is less than $\alpha$. Or, in hypothesis
testing lingo: if $\theta$ were one of the interval boundaries, then
everything beyond $\hat{\theta}$ would fall into the rejection region.
When the probability $\alpha$ is distributed evenly among small and large
deviations, the formal definition of the {\em frequentist confidence
interval} becomes\footnote{This definition reads slightly different from the definition given by DiCiccio \& Efron \cite{diciccio96}: Eq.~(\ref{eq:CIfreq:b}) is identical, but in Eq.~(\ref{eq:CIfreq:a}) they write ``$>$'' instead of ``$\geq$''. This makes no difference for continuous random variables, but it would treat the two boundaries differently for discrete random variables.}:
\begin{subequations}
\label{eq:CIfreq}
\begin{eqnarray}
\label{eq:CIfreq:a}
P_{\theta=\theta_l}(\hat{\theta}\geq\theta_0) & = & \alpha/2 \quad\mbox{and}\\
\label{eq:CIfreq:b}
P_{\theta=\theta_u}(\hat{\theta}\leq\theta_0) & = & \alpha/2
\end{eqnarray}
\end{subequations}
where $\theta_0$ is the observed value for the estimator
and $P_{\theta=\theta_{l,u}}$ is the probability under the assumption
that the true parameter value is the lower or upper boundary, respectively.

%-----------------------------------------------------------------------
\begin{figure}[t]
\includegraphics[width=\columnwidth]{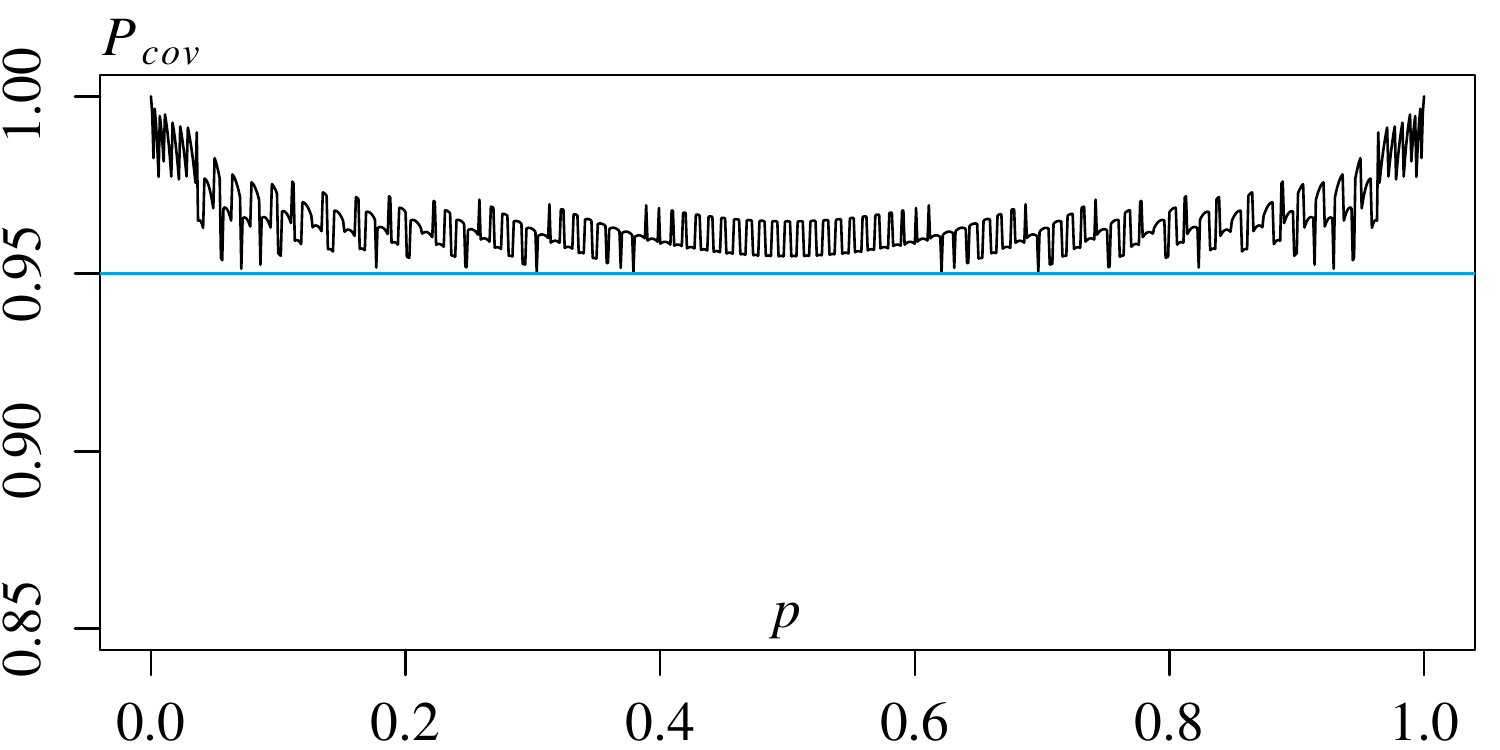}
\caption{\label{fig:pcov-binom-exact}Coverage probability $P_{cov}$ of the ``exact'' confidence interval for a binomial proportion after Eq.~(\ref{eq:CIfreq}) as a function of the true parameter $p$ for $n=100$ and $\alpha=0.05$.}
\end{figure}
%-----------------------------------------------------------------------

Although the confidence interval obtained by solving Eq.~(\ref{eq:CIfreq})
for $\theta_l$ and $\theta_u$ is guaranteed to have have at least
$1-\alpha$ coverage probability independent from $\theta$, there are
two hitches: the example in Fig.~\ref{fig:pcov-binom-exact} shows that
even an ``exact'' confidence interval directly computed with
Eq.~(\ref{eq:CIfreq}) can have coverage probability that is too large
for most values of $\theta$, which means that the interval is too wide.
Moreover, the probability is often known only approximately, or
Eq.~(\ref{eq:CIfreq}) can only be solved asymptotically, which leads
to an approximate confidence interval, which can have $P_{cov}(\theta)$
less than $1-\alpha$.

\subsection{Likelihood ratio}
%-----------------------------------------------------------------------
\label{sec:grundbegriffe:lr}
A different approach to obtain a confidence interval is based on the
likelihood function (\ref{eq:L:a}). The ML estimator $\hat{\theta}$
chooses $\theta$ such that it maximizes the probability of the observed data.
However, other values of $\theta$ lead to a high probability of the observation,
too. It is thus natural to define an interval wherein the ratio
$L(\hat{\theta})/L(\theta)$ is greater than some threshold. To distinguish
this interval from the frequentist confidence interval, it is called the
{\em likelihood ratio support interval} $[\theta_l,\theta_u]$:
\begin{equation}
\label{eq:LR}
\frac{L(\theta)}{L(\hat{\theta})} \geq \frac{1}{K} \quad\mbox{for all }
\theta\in [\theta_l,\theta_u]
\end{equation}
where $\hat{\theta}$ is the ML estimator for $\theta$. A common choice for
$K$ is $K=8$ because, in the case of mean values, it leads to intervals
very close to the frequentist interval for $\alpha=0.05$
(see section \ref{sec:mu:lr}).

\subsection{Posterior density}
\begin{figure}[t]
\centering\includegraphics[width=0.9\columnwidth]{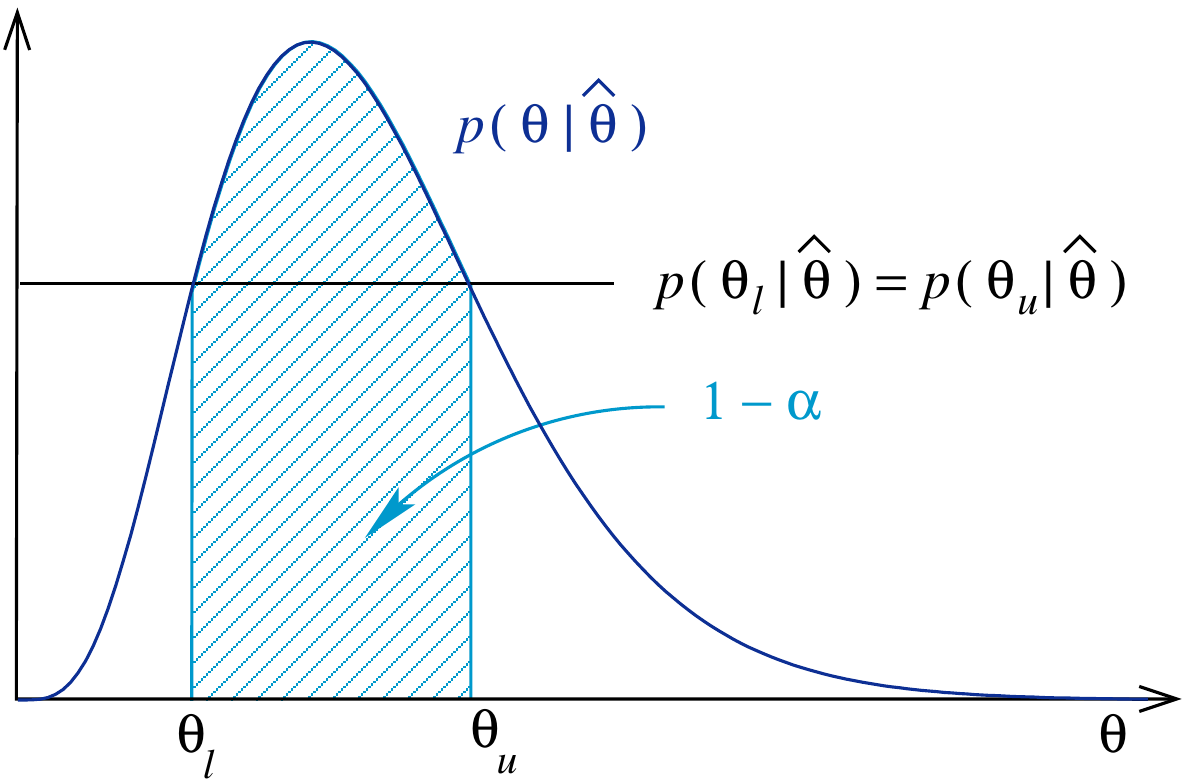}
\caption{\label{fig:hpd}Determination of the highest posterior density interval  $[\theta_l,\theta_u]$ according to Eq.~(\ref{eq:hpd}).}
\end{figure}

%-----------------------------------------------------------------------
\label{sec:grundbegriffe:hpd}
A third approach to confidence interval construction tries to estimate
a probability density for $\theta$ on basis of the observation $\hat{\theta}$.
The true parameter $\theta$ is here considered as a random variable, and
$p_\theta(\hat{\theta})$ is a conditional probability density\footnote{Note that $\theta$ and $\hat{\theta}$ are continuous variables, so that their probability distribution is described by a density, here denoted with the lower case letter $p$.}
$p(\hat{\theta}|\theta)$ that can be computed with Bayes' formula:
\begin{equation}
\label{eq:bayes}
p(\theta|\hat{\theta}) = \frac{p(\hat{\theta}|\theta)\cdot p(\theta)}{\int_{\mathbb{R}}p(\hat{\theta}|\tau)\cdot p(\tau)\,d\tau}
\end{equation}
Based on this density, the {\em highest posterior density (HPD) interval}
is defined as the region $[\theta_l,\theta_u]$ with highest probability
density values and a total probability of $(1-\alpha)$. Formally, this definition
leads to the coupled equations (see Fig.~\ref{fig:hpd})
\begin{subequations}
\label{eq:hpd}
\begin{eqnarray}
1-\alpha &=& \int_{\theta_l}^{\theta_u}\!\! p(\theta|\hat{\theta})\,d\theta 
\quad\quad\mbox{and}\\
p(\theta_l|\hat{\theta}) &=& p(\theta_u|\hat{\theta})
\end{eqnarray}
\end{subequations}

%-----------------------------------------------------------------------
%-----------------------------------------------------------------------

Apart from the nuisance that this system of equations can only be solved
numerically, the HPD interval has a fundamental deficiency: to compute
$p(\theta|\hat{\theta})$ with Eq..~(\ref{eq:bayes}), it is necessary
to make an assumption about the ``a priori distribution'' $p(\theta)$
of the unknown parameter $\theta$, and this assumption is arbitrary.
Typically, $p(\theta)$ is chosen to be constant which implies that
nothing is known about the approximate location of $\theta$. Although
this assumption is rarely realistic in practical situations, this does
not necessarily mean that the HPD interval is bad. As we will see in the
next section, it can even have a good coverage probability.

\section{Relative frequencies}
%-----------------------------------------------------------------------
\label{sec:p}
The relative frequency $\hat{p}$ is a ubiquitous estimator for a probability,
or a binomial proportion $p$. The probability distribution of $\hat{p}$
is exactly given by the binomial distribution. When an event has probability
$p$, the probability that it occurs $k$ times in $n$ independent trials is
\begin{equation}
\label{eq:binom}
P_p(k) = {n \choose k} p^k (1-p)^{n-k}
\end{equation}
The relative frequency $\hat{p}=k/n$ then has the probability
\begin{equation}
\label{eq:binom}
P_p(\hat{p}=p_0) = {n \choose np_0} p^{np_0} (1-p)^{n(1-p_0)}
\end{equation}
Eq.~(\ref{eq:binom}) is the starting point for all confidence intervals of
the relative frequency.

\subsection{Frequentist interval for $\hat{p}$}
%-----------------------------------------------------------------------
\label{sec:p-mu:frequentistisch}

%------------------------------------------------------------------------- 
\lstset{language=R,
  basicstyle=\small \ttfamily,
  literate={.help}{.help}5,
  keywordstyle=\ttfamily,
  frame=bottomline,
  floatplacement=!t,
  aboveskip=0pt,
  belowskip=0pt,
  captionpos=b
}
\begin{lstlisting}[float, caption=R implementation of the exact Clopper-Pearson confidence interval for the relative frequency after Eqs.~(\ref{eq:freq:pexakt}) \& (\ref{eq:freq:pexakt:0n})., label=lst:freq:pexakt]
ci.binom <- function(n, k, alpha) {
  if (k == 0) {
    p1 <- 0.0
    p2 <- 1 - (alpha/2)**(1/n)
  }
  else if (k == n) {
    p1 <- (alpha/2)**(1/n)
    p2 <- 1.0
  }
  else {
    helper <- function(p, k, n, val) {
      return (pbinom(k, n, p) - val)
    }
    r <- uniroot(helper, k=(k-1),
                 n=n, val=1-alpha/2,
                 interval=c(0,1))
    p1 <- r$root
    r <- uniroot(helper, k=k,
                 n=n, val=alpha/2,
                 interval=c(0,1))
    p2 <- r$root
  }
  return (data.frame(p1=p1, p2=p2))
}
\end{lstlisting}
%------------------------------------------------------------------------- 

Insertion of (\ref{eq:binom}) into Eq.~(\ref{eq:CIfreq}) yields the 
following equations to determine boundaries $p_l$ and $p_u$:
\begin{subequations}
\label{eq:freq:pexakt}
\begin{eqnarray}
&  1-\mbox{pbinom}\left((k-1)/n,n,p_l\right) = \alpha/2 & \\
\mbox{and} & 
\mbox{pbinom}\left(k/n,n,p_u\right) = \alpha/2 &
\end{eqnarray}
\end{subequations}
where $k/n=\hat{p}$ is the observed relative frequency, and {\em pbinom}
is the R function for the cumulative distribution function (CDF) of
the binomial distribution. In the special cases $k=0$ or $k=0$, one of
the equations (\ref{eq:freq:pexakt}) does not have a solution because
$p_l$ and $p_u$ are restricted to the interval $[0,1]$. In these cases,
let $p_l=0$ ($k=0$) or $p_u=1$ ($k=n$), respectively. The other boundary
can be found analytically in these cases as
\begin{subequations}
\label{eq:freq:pexakt:0n}
\begin{eqnarray}
k=0 & \Rightarrow & [p_l,p_u] = [0, 1-\sqrt[n]{\alpha/2}] \\
k=n & \Rightarrow & [p_l,p_u] = [\sqrt[n]{\alpha/2}, 1]
\end{eqnarray}
\end{subequations}
In all other cases, Eq.~(\ref{eq:freq:pexakt}) must be solved numerically,
e.g., with the R function {\em uniroot}\footnote{It would also be possible to use the inverse of the incomplete beta function, because 1-{\em pbinom} can be expressed through this function (see \cite{abramowitz84} Eq.~26.5.7). The inverse of the incomplete beta function, however, must be computed numerically either.}.
The corresponding R code is given in listing \ref{lst:freq:pexakt}.
This interval is known as the {\em Clopper-Pearson} interval \cite{clopper34},
which is also implemented in the R function {\em binom.confint} from the
package {\em binom}, with the option {\em method='exact'}.

An approximate confidence interval is obtained when the binomial distribution
is replaced by the normal distribution, which is justified by the central limit
theorem. For large $n$, $\hat{p}$ is approximately normally distributed with
$\mu=p$ and $\sigma^2=p(1-p)/n$. With this approximation, 
Eq.~(\ref{eq:CIfreq:a}) becomes
\begin{align}
& 1-\mbox{pnorm}\left(\hat{p},p_l,\sqrt{p_l(1-p_l)/n}\right) = \alpha/2 \nonumber \\
\Leftrightarrow\quad & \mbox{pnorm}\left(\frac{\hat{p}-p_l}{\sqrt{p_l(1-p_l)/n}},0,1\right) = 1-\alpha/2 \nonumber \\
\Leftrightarrow\quad & \frac{\hat{p}-p_l}{\sqrt{p_l(1-p_l)/n}} = z_{1-\alpha/2}
\label{eq:wilson:ansatz}
\end{align}
where {\em pnorm} is the R function for the CDF of the normal distribution,
and $z_{1-\alpha/2}=\mbox{qnorm}(1-\alpha/2)$ is the $(1-\alpha/2)$ quantile
of the standard normal distribution. The quadratic equation
(\ref{eq:wilson:ansatz}) and its analogous version for $p_u$ can be solved
analytically, thereby yielding the {\em Wilson interval}:
\begin{equation}
\label{eq:freq:pwilson}
\frac{1}{1+z^2/n}\left[\hat{p} + \frac{z^2}{2n} \pm z \sqrt{\frac{\hat{p}(1-\hat{p})}{n}+\frac{z^2}{4n^2}}\right]
\end{equation}
where $z=z_{1-\alpha/2}$, for the sake of brevity. In the comparative study
\cite{brown01}, Brown et al.~recommended the Wilson interval due to its
coverage probability. For large $n$, Eq.~(\ref{eq:freq:pwilson}) asymptotically
transforms into the classical {\em Wald interval} that is taught in
introductory text books:
\begin{equation}
\label{eq:freq:pwald}
\hat{p} \pm z_{1-\alpha/2} \sqrt{\hat{p}(1-\hat{p})/n}
\end{equation}

%-----------------------------------------------------------------------
\begin{figure}[t]
\centering\includegraphics[width=0.9\columnwidth]{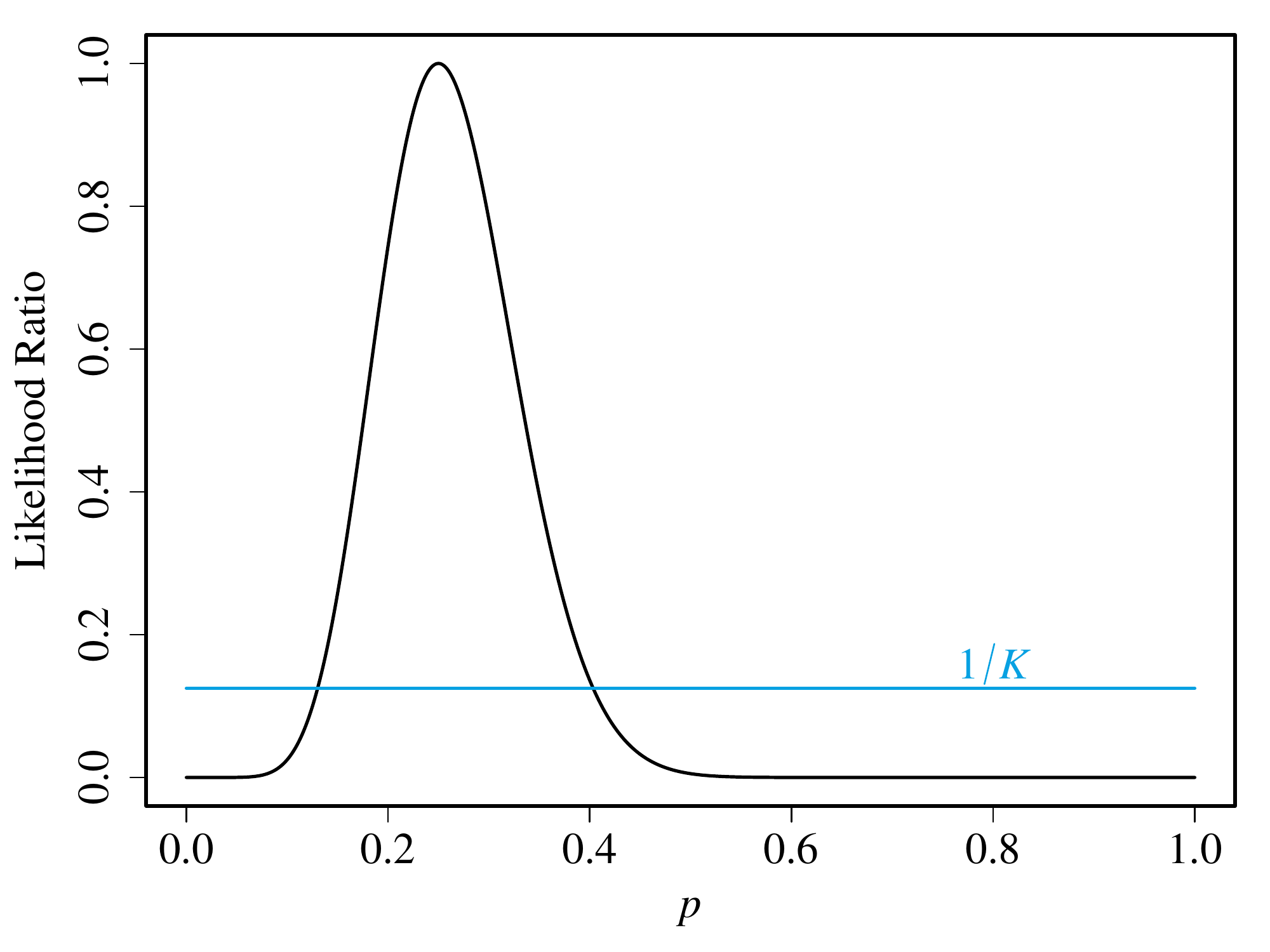}
\caption{\label{fig:lr-binom}Likelihood ratio $L(p)/L(\hat{p})$ of the binomial distribution for $n=40$ and $k=10$.}
\end{figure}
%-----------------------------------------------------------------------

\subsection{Likelihood ratio for $\hat{p}$}
%-----------------------------------------------------------------------
\label{sec:p:lr}
When the event of interest occurs $k$ times in $n$ trials, the likelihood
function is
\begin{equation}
L(p) = p^k(1-p)^{n-k}
\end{equation}
The relative frequency $\hat{p}=k/n$ is the ML estimator for $p$. The
likelihood ratio support interval therefore encompasses all $p$ with
\begin{equation}
\label{eq:lr:binom}
\frac{L(p)}{L(\hat{p})} = \frac{p^k(1-p)^{n-k}}{\hat{p}^k(1-\hat{p})^{n-k}}
\geq \frac{1}{K}
\end{equation}
A plot of the function on the left hand side is shown in
Fig.~\ref{fig:lr-binom}. Eq.~(\ref{eq:lr:binom}) must be solved numerically,
e.g., with the R function {\em uniroot}. A possible implementation is given
in listing \ref{lst:lr-binom}.

%------------------------------------------------------------------------- 
\lstset{language=R,
  basicstyle=\small \ttfamily,
  literate={.help}{.help}5,
  keywordstyle=\ttfamily,
  frame=bottomline,
  floatplacement=!t,
  aboveskip=0pt,
  belowskip=0pt,
  captionpos=b
}
\begin{lstlisting}[float, caption=R code that computes the likelihood ratio support interval for the relative frequency according to Eq.~(\ref{eq:lr:binom})., label=lst:lr-binom]
lr.binom <- function(n, k, K) {
  helper <- function(p, n, k, K) {
    return (p**k * (1-p)**(n-k) / 
            ((k/n)**k * (1-k/n)**(n-k))
            - 1/K)
  }
  p1 <- rep(0,length(k))
  p2 <- p1
  if (k==0) {
    p1 <- 0
  } else {
    r <- uniroot(helper,n=n,k=k,K=K,
                 interval=c(0,k/n))
    p1 <- r$root
  }
  if (k==n) {
    p2 <- 1
  } else {
    r <- uniroot(helper,n=n,k=k,K=K,
                 interval=c(k/n,1))
    p2 <- r$root
  }
  return (data.frame(p1=p1, p2=p2))
}
\end{lstlisting}
%------------------------------------------------------------------------- 

\subsection{Highest posterior density for $\hat{p}$}
%-----------------------------------------------------------------------
\label{sec:p:hpd}
The R package {\em HDInterval} provides the function {\em hdi} for
computation of HPD intervals. {\em hdi} requires as one
function argument a function that computes the inverse of
$\int_{-\infty}^{\theta} p(\tau|\hat{\theta})\,d\tau$. This means that
{\em hdi} is only applicable in cases where this inverse function can
be readily computed. The binomial distribution is such a case.

Insertion of the binomial distribution (\ref{eq:binom}) into 
Eq.~(\ref{eq:bayes}) yields with the assumption of a constant
``a priori'' density $p(\theta)=\mbox{\em const.}$:
\begin{eqnarray}
p(p|k) & = &
\frac{ {n \choose k} p^k (1-p)^{n-k} }{\int_{0}^{1} {n \choose k} q^k (1-q)^{n-k}\,dq} \nonumber \\[1ex]
 & = & \frac{\Gamma(a+b)}{\Gamma(a)\Gamma(b)}\,p^{a-1}(1-p)^{b-1} \nonumber \\[1ex]
& = & \mbox{dbeta}(p, a, b)
\end{eqnarray}
where $a=k+1$ and $b=n-k+1$, and {\em dbeta} is the R function for the
probability density of the beta distribution. The inverse CDF of the
beta distribution in provided by R as the function {\em qbeta}, so that
the HPD interval can be computed with the code in listing \ref{lst:hpd-binom}.

%------------------------------------------------------------------------- 
\lstset{language=R,
  basicstyle=\small \ttfamily,
  keywordstyle=\ttfamily,
  frame=bottomline,
  floatplacement=!t,
  aboveskip=0pt,
  belowskip=0pt,
  captionpos=b
}
\begin{lstlisting}[float, caption=R code that computes the $(1-\alpha)$ HPD interval for the relative frequency., label=lst:hpd-binom]
library(HDInterval)
ci <- hdi(qbeta, 1-alpha,
          shape1=(k+1),
          shape2=(n-k+1))
p1 <- ci[1]; p2 <- ci[2]
\end{lstlisting}
%------------------------------------------------------------------------- 

\section{Mean values}
%-----------------------------------------------------------------------
\label{sec:mu}
Another ubiquitous estimator is the statistical average $\overline{x}$
as an estimator for the expectation value $\mu=E(X)$. For the statistical
average $\overline{x}=\frac{1}{n}\sum_{i=1}^nx_i$, it is possible to construct
a quantity that only depends on the unknown $\mu$ and has a known distribution,
albeit only in the special case that the variable $X$ is normally distributed.
In this case, the random variable
\begin{equation}
\label{eq:xquer-standard}
Z=\frac{\overline{x}-\mu}{\sqrt{s^2/n}}\quad\mbox{with }
s^2=\frac{1}{n-1}\sum_{i=1}^n (x_i-\overline{x})^2
\end{equation}
is $t$ distributed with $(n-1)$ degrees of freedom\footnote{The esoterically sounding term ``degrees of freedom'' is just the parameter of the $t$ distribution.}.
If $X$ is not normally distributed, it is at least known from the central limit
theorem that the quantity (\ref{eq:xquer-standard}) is approximately
standard normally distributed\footnote{The ``standard'' normal distribution is the normal distribution with parameters $\mu=0$ and $\sigma^2=1$.}
\cite{fahrmeir04}. In general, it is not known whether $X$ is normally
distributed, which means that confidence intervals for the mean value can
alternatively be based on the $t$ distribution or the normal distribution.

\subsection{Frequentist interval for $\mu$}
%-----------------------------------------------------------------------
\label{sec:mu:frequentistisch}
Let $\mu_0$ be the observed value for $\overline{x}$. Then the 
Eq.~(\ref{eq:CIfreq:a}) specifying $\mu_l$ reads, with utilization of the
$t$ distribution:
\begin{align}
\label{eq:freq:mu-ansatz}
 &P_{\mu=\mu_l}(\overline{x}\geq \mu_0) = \alpha/2 \\ \nonumber
\Leftrightarrow\quad &P\left(Z\geq (\mu_0-\mu_l)/\sqrt{s^2/n}\right) = \alpha/2 \\ \nonumber
\Leftrightarrow\quad & 1- \mbox{pt}\left((\mu_0-\mu_l)/\sqrt{s^2/n}, n-1\right) = \alpha/2 \\ \nonumber
\Leftrightarrow\quad & (\mu_0-\mu_l)/\sqrt{s^2/n} = \mbox{qt}(1-\alpha/2, n-1) \\ \nonumber
\Leftrightarrow\quad & \mu_l = \mu_0-\mbox{qt}(1-\alpha/2, n-1)\cdot \sqrt{s^2/n}
\end{align}
where {\em pt} is the CDF of the $t$ distribution, and {\em qt} its inverse.
In the same way, Eq.~(\ref{eq:CIfreq:b}) can be solved for $\mu_u$.
With utilization of the symmetry property $\mbox{qt}(t) = -\mbox{qt}(1-t)$,
the confidence interval based upon the $t$ distribution becomes:
\begin{equation}
\label{eq:freq:mu-t}
\overline{x}\pm t_{1-\alpha/2}(n-1)\cdot\sqrt{s^2/n}
\end{equation}
where $t_{1-\alpha/2}(n-1)$ denotes the $(1-\alpha/2)$ quantile of the $t$
distribution, which can be computed with the R function {\em qt}.

Based on the normal distribution, the same calculation method yields the 
confidence interval
\begin{equation}
\label{eq:freq:mu-z}
\overline{x}\pm z_{1-\alpha/2}\cdot\sqrt{s^2/n}
\end{equation}
where $z_{1-\alpha/2}$ denotes the $(1-\alpha/2)$ quantile of the 
standard normal distribution, which can be computed with the R function
{\em qnorm}.

It seems paradoxical that we obtain the different confidence intervals
(\ref{eq:freq:mu-t}) or (\ref{eq:freq:mu-z}), depending on a condition
(the underlying distribution) that we do not know about. This is 
no contradiction, however. Although
\begin{equation}
t_{1-\alpha/2}(n-1) > z_{1-\alpha/2} \quad\mbox{for all }n
\end{equation}
and the interval (\ref{eq:freq:mu-t}) is therefore always slightly larger,
for large $n$ both intervals become asymptotically similar because of
\begin{equation}
\lim_{n\to\infty} t_{1-\alpha/2}(n-1) = z_{1-\alpha/2}
\end{equation}
For $\alpha=0.05$, both values are close to two, which leads for both of the
above confidence intervals to the rule of thumb ``two times sigma'', with
$\sigma=\sqrt{s^2/n}$.

\subsection{Likelihood ratio for $\mu$}
%-----------------------------------------------------------------------
\label{sec:mu:lr}
On basis of the $t$ distribution the specifying equation (\ref{eq:LR})
for the likelihood ratio support interval reads
\begin{equation}
\frac{L(\mu)}{L(\hat{\mu})} = \left(1+\frac{n(\overline{x}-\mu)^2}{s^2(n-1)}\right)^{-n/2} \geq \frac{1}{K}
\end{equation}
This equation can readily be solved for $\mu$, which yields the support interval
\begin{equation}
\label{eq:mu:lr-t}
\overline{x}\pm\sqrt{(K^{2/n}-1) s^2\frac{n-1}{n}}
\end{equation}
On basis of the normal distribution the specifying equation reads
\begin{equation}
%\frac{L(\mu)}{L(\hat{\mu})} = \left(1+\frac{n(\overline{x}-\mu)^2}{s^2(n-1)}\right)^{-n/2} \geq \frac{1}{K}
\frac{L(\mu)}{L(\hat{\mu})} = \exp\left(-\frac{n(\overline{x}-\mu)^2}{2s^2}\right) \geq \frac{1}{K}
\end{equation}
This can again be solved elementary for $\mu$, too, which yields the support 
interval
\begin{equation}
\label{eq:mu:lr-z}
\overline{x}\pm\sqrt{\frac{2s^2}{n}\ln K}
\end{equation}
It seems as though (\ref{eq:mu:lr-t}) and (\ref{eq:mu:lr-z}) were completely
different intervals, but in fact they are very similar: for large $n$, both
intervals are asymptotically equal because of\footnote{This limiting value follows from inversion of \cite{abramowitz84} Eq.~4.2.21.}
\begin{equation}
\ln x = \lim_{n\to\infty} n(x^{1/n} - 1)
\end{equation}
The numerical evaluation of the right hand side of Eq.~(\ref{eq:mu:lr-t})
becomes inaccurate, however, for large $n$ due to extinction of the most leading
digit from from similar floating point numbers. Therefore, 
Eq.~(\ref{eq:mu:lr-z}) is preferable for large $n$ even in the case of the
$t$ distribution.

When we compare the support interval (\ref{eq:mu:lr-z}) with the
confidence interval (\ref{eq:freq:mu-t}), we see the reason for the choice
$K=8$: it is $\sqrt{2\ln 8}\approx 2.0393$, which means that the frequentist
interval for $\alpha=0.05$ and the LR support interval roughly coincide.
For $K=7$, it is even with good accuracy $\sqrt{2\ln K}\approx z_{1-\alpha/2}$,
but, as we will see in section \ref{sec:vergleich}, the frequentist interval
based on $z_{1-\alpha/2}$ is generally to small, so that $K=8$ is a safer choice.

%% %------------------------------------------------------------------------- 
%% \lstset{language=R,
%%   basicstyle=\small \ttfamily,
%%   literate={.t}{.t}2,
%%   keywordstyle=\ttfamily,
%%   frame=bottomline,
%%   floatplacement=!t,
%%   aboveskip=0pt,
%%   belowskip=0pt,
%%   captionpos=b
%% }
%% \begin{lstlisting}[float, caption=R Implementierung des HPD-Intervalls für den statistischen Mittelwert. {\em xquer} ist der gemessene Mittelwert und {\em s} die gemessene Streuung., label=lst:mu-hpd]
%% library(HDInterval)

%% # basierend auf der t-Verteilung
%% Finv.t <- function(p, df, mult, plus) {
%%   return(qt(p, df=df)*mult + plus)
%% }
%% ci <- hdi(Finv.t, 1-alpha, df=(n-1),
%%           mult=s/sqrt(n), plus=xquer)
%% mu1 <- ci[1]; mu2 <- ci[2]

%% # basierend auf der Normalverteilung
%% ci <- hdi(qnorm, 1-alpha,
%%           mean=xquer, sd=s/sqrt(n))
%% mu1 <- ci[1]; mu2 <- ci[2]
%% \end{lstlisting}
%% %------------------------------------------------------------------------- 

\subsection{Highest posterior density for $\mu$}
%-----------------------------------------------------------------------
\label{sec:mu:hpd}
On basis of the $t$ distribution, Eq.~(\ref{eq:bayes}) becomes with the
assumption of a constant ``a priori'' distribution $p(\mu)=\mbox{\em const.}$:
\begin{align}
p(\mu|\overline{x}) & =  \frac{\sqrt{n}\Gamma(\frac{n}{2})}{s\sqrt{\pi(n-1)}\Gamma(\frac{n-1}{2})}\left(1+\frac{(\overline{x}-\mu)^2n}{s^2(n-1)}\right)^{\!\!-\frac{n}{2}} \nonumber \\[1ex]
 & =   \sqrt{\frac{n}{s^2}}\cdot\mbox{dt}\!\left(\frac{(\overline{x}-\mu)\sqrt{n}}{s}, n-1\right)
\end{align}
where {\em dt} is the R function for the probability density of the $t$
distribution. On basis of the normal distribution, we obtain under the
analogous assumption $p(\mu)=\mbox{\em const.}$:
\begin{align}
p(\mu|\overline{x}) &=
      \sqrt{\frac{n}{2\pi s^2}}\cdot\exp\left(-\frac{(\overline{x}-\mu)^2n}{2s^2}\right) \nonumber \\[1ex]
 &= \mbox{dnorm}(\mu, \overline{x}, s^2\!/n)
\end{align}
where {\em dnorm} is the R function for the probability density of the
normal distribution. The resulting densities are thus identical to
the symmetric densities used for the frequentist interval, which has the
effect that the specifying equation (\ref{eq:freq:mu-ansatz}) for the HPD
interval has the same solution as the specifying equation for the 
frequentist interval. The HPD interval for the mean value is therefore
exactly identical to the frequentist interval (\ref{eq:freq:mu-t})
or (\ref{eq:freq:mu-t}), respectively.

This is no coincidence, but a consequence of the fact that $\mu$ is a
``location parameter'', i.e., that $p(\overline{x}|\mu)=f(\overline{x}-\mu)$.
When this functional relationship holds, frequentist interval and HPD interval
are always identical \cite{karlen02}.

\section{Maximum likelihood estimators}
%-----------------------------------------------------------------------
\label{sec:ml}
To obtain a confidence interval for different estimators, it is necessary
to know the probability distribution of the estimated value $\hat{\theta}$.
Unfortunately, this is almost impossible in other cases than the aforementioned
two examples. There is however a large category of estimators for which
the asymptotic distribution is known: maximum likelihood (ML) estimators
are asymptotically normally distributed around the true value $\theta$
for ``regular'' log-likelihood functions\footnote{\label{fn:ml-normal}The precise requirements are as follows: the log-likelihood function $\ell(\theta)$ must be three times continuously differentiable, the expectation values of all first and second derivatives exist, and the third derivations must be bounded by a function with finite expectation value \cite{greene00}.}
$\ell(\theta)$ (see Eq.~(\ref{eq:L:b})) for large $n$. In other words,
the asymptotic probability density of $\hat{\theta}$ is given by
\begin{equation}
\label{eq:normalverteilung}
p(\hat{\theta}) = \frac{\exp\left(-\frac{1}{2}\langle\hat{\theta}-\theta, \Sigma^{-1} (\hat{\theta}-\theta)\rangle\right)}{\sqrt{(2\pi)^t\det(\Sigma)}}
\end{equation}
where $t$ is the number of parameters $\theta=(\theta_1,\ldots,\theta_t)$, 
$\Sigma$ is the covariance matrix, and the exponent ``$-1$'' denotes matrix 
inversion.

If it is thus possible to determine the covariance matrix
$(\sigma_{ij})=\Sigma$, then its diagonal elements 
$\sigma_{ii}=\mbox{\em Var}(\theta_i)$ can be used to construct confidence
intervals based on the normal distribution as in section \ref{sec:mu}:
\begin{equation}
\label{eq:ml-klassisch}
\hat{\theta} \pm z_{1-\alpha/2}\sqrt{\sigma_{ii}}
\end{equation}
Alternatively, it would also be sufficient to have a direct estimator
for the variances $\sigma_{ii}$ of the parameters. This leads to two
possible approaches for an estimation of the variance of maximum likelihood
estimators:
\begin{itemize}
\item estimation of the covariance matrix via inversion of the Hessian matrix
  of the log-likelihood function
\item jackknife estimator for the variance
\end{itemize}
The first method has the advantage that it can yield closed formulas for
the variance in cases that allow for an analytic calculation of the Hessian
matrix. The second method has the advantage that it requires no analytic or
numeric calculation of derivatives at all, but that it provides an elementary 
and fast algorithm for computing the variance.

When the requirements listed in footnote \ref{fn:ml-normal} do not hold,
the Hessian matrix cannot be computed, and the first method is ruled out.
Although the jackknife variance can nevertheless be computed even in this case,
it is of little use, because neither is guaranteed that the estimator is
normally distributed, nor that the jackknife variance is a good estimator
for the true variance (see \cite{miller64} for a counterexample).
In such a situation, it is therefore necessary to resort to the
bootstrap method which is described in section
\ref{sec:bootstrap}.

\subsection{Hessian matrix}
%-----------------------------------------------------------------------
\label{sec:ml:hesse}
When the preconditions mentioned in footnote \ref{fn:ml-normal} hold,
the covariance matrix in Eq.~(\ref{eq:normalverteilung}) can be estimated
through
\cite{greene00}
\begin{equation}
\label{eq:ml-hesse}
\left(\sigma_{ij}\right) = \left(-\left.\frac{\partial^2 \ell}{\partial\theta_i \partial\theta_j}\right|_{\theta=\hat{\theta}}\right)^{-1}
\end{equation}
where $\ell(\theta)$ is the log-likelihood function form Eq.~(\ref{eq:L:b}),
and the exponent ``$-1$'' denotes matrix inversion.

%------------------------------------------------------------------------- 
\lstset{language=R,
  basicstyle=\small \ttfamily,
  literate={.help}{.help}5,
  keywordstyle=\ttfamily,
  frame=bottomline,
  floatplacement=!t,
  aboveskip=0pt,
  belowskip=0pt,
  captionpos=b
}
\begin{lstlisting}[float, caption={R code for the numerical calculation of an
ML estimator for $\theta=(\theta_1,\ldots,\theta_t)$ in combination with a
variance estimation for the estimated values. The log-likelihood function
must be defined negatively, because {\em optim} seeks the minimum instead
of the maximum.}, label=lst:ml]
lnL <- function(theta1, theta2, ...) {
  # definition of the negative (!)
  # log-likelihood function
  ...
}

# starting values for the optimization
theta0 <- c(start1, start2, ...)

# optimization
p <- optim(theta0, lnL, hessian=TRUE)
if (p$convergence == 0) {
  theta <- p$par
  covmat <- solve(p$hessian)
  sigma <- sqrt(diag(covmat))
}
\end{lstlisting}
%------------------------------------------------------------------------- 

%$ (fix for highlighting in emacs latex-mode)

In many cases, neither the equation (\ref{eq:ML}) specifying the ML
estimator $\hat{\theta}$ can be solved in closed form, nor can the inverse
of the Hessian matrix (\ref{eq:ml-hesse}) be computed analytically.
This does not mean, however, that this method must be ruled out in this case,
because a numerical solution is often viable. The R function {\em optim}
even offers an argument {\em hessian=TRUE} which asks for an additional
estimation of the Hessian matrix during optimization. An example implementation
utilizing this function is given in listing \ref{lst:ml}.

\subsection{Jackknife}
%-----------------------------------------------------------------------
\label{sec:ml:hesse}
The jackknife method is based on the idea to compute the estimator 
$\hat{\theta}(x_1,\ldots,x_n)$ many times, but each time with the omission of
one value $x_i$. The variance of $\hat{\theta}$ is then estimated form the
distribution of these ``delete-one'' estimators. Let $\theta_{(i)}$ be the
estimator computed without the $i$-th data point $x_i$. Then the jackknife
estimator for the variance of $\hat{\theta}$ is:
\begin{align}
\label{eq:sigma-jk}
\sigma_{\mbox{\scriptsize\it JK}}(\hat{\theta}) &= \sqrt{\frac{n-1}{n}\sum_{i=1}^n (\theta_{(i)}-\theta_{(.)})^2}\\
 \mbox{ with }&\quad \theta_{(.)}=\frac{1}{n}\sum_{i=1}^n\theta_{(i)} \nonumber
\end{align}
When $\theta$ is a vector with several components, it is also possible to
estimate the entire covariance matrix $Sigma$ with the jackknife. This is
of little use however, because the confidence intervals (\ref{eq:ml-klassisch})
only need the diagonal elements $\sigma_{ii}$ of $\Sigma$. Hence it is
sufficient to apply Eq.~(\ref{eq:sigma-jk}) to each component of 
$\theta$. For asymptotically normally distributed ML estimators,
$\sigma_{\mbox{\scriptsize\it JK}}$ is an asymptotically unbiased and consistent
estimator for their variance \cite{reeds78}. An implementation of
formula (\ref{eq:sigma-jk}) is given in listing \ref{lst:jackknife}.

%------------------------------------------------------------------------- 
\lstset{language=R,
  basicstyle=\small \ttfamily,
  literate={ü}{{\"u}}1{ä}{{\"a}}1,
  keywordstyle=\ttfamily,
  frame=bottomline,
  floatplacement=!t,
  aboveskip=0pt,
  belowskip=0pt,
  captionpos=b
}
\begin{lstlisting}[float, caption={Calculation of the jackknife variance of an estimator $\hat{\theta}(x_1,\ldots,x_n)$ in R.}, label=lst:jackknife]
theta.hat <- function(x) {
  # implementation of the estimator
  ...
}
theta.jk <- rep(0, n)
for (i in 1:n) {
  theta.jk[i] <- theta.hat(x[-i])
}
theta.dot <- mean(theta.jk)
sigma.jk <- sqrt((n-1) *
        mean((theta.jk-theta.dot)^2))
\end{lstlisting}
%------------------------------------------------------------------------- 

\section{Bootstrap}
%-----------------------------------------------------------------------
\label{sec:bootstrap}
Similar to the jackknife method, the bootstrap method is based on the
generation of new data sets from the original data $x_1,\ldots,x_n$, In the
bootstrap, this is however not done deterministically via cyclic omission,
but in a random way. This can either be be done by $n$-fold drawing
with replacement ({\em non-parametric bootstrap}), or by $n$-fold generation
of random numbers distributed according to the density estimated with the
estimator ({\em parametric bootstrap}). The non-parametric bootstrap thus
considers all observed data, while the parametric bootstrap only considers
the point estimator $\hat{\theta}$ computed from the data.

When we repeat the drawing of new data sets $R$ times, we obtain a Monte-Carlo
simulation of the distribution of the estimator $\hat{\theta}$. From this
distribution, confidence intervals can be estimated.\footnote{It is also possible to estimate the variance from this Monte Carlo simulation \cite{efron83}, but a confidence interval based on this variance would again make the assumption of a normally distributed $\hat{\theta}$.}\footnote{You could think that instead of the bootstrap random samples, one could alternatively estimate the confidence interval from the distribution of the $n$ jackknife ``delete one'' estimators $\theta_{(i)}$. This does not work, however, because even in the case of regular ML estimators, the distribution of the $\theta_{(i)}$ is not normal and therefore not representative for the distribution of $\hat{\theta}$ \cite{wu86}.}.
There is a bewildering variety of methods for estimating a confidence interval
from the simulated distribution, which are summarized together with their
asymptotic coverage probability in \cite{carpenter00}. Their theoretical 
background is explained in \cite{diciccio96}. The most important methods are:
\begin{description}
\item[Percentile \& Basic.] The {\em Percentile bootstrap} was the original
  method proposed by Efron. It simply takes the percentiles of the simulated
  distribution $\hat{\theta}_1,\ldots,\hat{\theta}_n$ of 
  $\hat{\theta}$. The {\em Basic bootstrap interval} flips the percentile
  bootstrap at $\hat{\theta}$. Venables \& Ripley recommend the Basic
  bootstrap over the Percentile bootstrap \cite{venables02}, but the experiments
  in section \ref{sec:vergleich} lead to the opposite conclusion.
\item[Bias corrected accelerated ($BC_a$).] This method tries to estimate
  transformation parameters that make the distribution symmetric.
  This is the method recommended by Efron.
\end{description}
It can be shown that the $BC_a$ interval has a coverage probability that
converges asymptotically for large $n$ to the nominal value $1-\alpha$ with
a rate $o(n^{-1})$ \cite{diciccio96}. This is a faster convergence than
for the classical $z_{1-\alpha/2}\sigma$ interval, which has a convergence
rate of $o(n^{-1/2})$. DiCiccio \& Efron concluded form this observation that
the bootstrap method is generally preferable (comments to \cite{diciccio96}, p.~228):
\begin{quote}
  ``If the standard intervals were invented today, they might not be
  publishable.''
\end{quote}

%------------------------------------------------------------------------- 
\lstset{language=R,
  basicstyle=\small \ttfamily,
  literate={ü}{{\"u}}1{ä}{{\"a}}1,
  keywordstyle=\ttfamily,
  frame=bottomline,
  floatplacement=!t,
  aboveskip=0pt,
  belowskip=0pt,
  captionpos=b
}
\begin{lstlisting}[float, caption={Calculation of bootstrap confidence intervals with the R library {\em boot}.}, label=lst:bootstrap]
# estimator function; indices are
# for boot() to select data points
schaetzer <- function(x, indices) {
  x.auswahl <- x[indices]
  ... # compute estimator from x.auswahl
  return(theta.hat)
}

# bootstrap confidence intervals
boot.out <- boot(data=x,
        statistic=schaetzer, R=1000)
ci <- boot.ci(boot.out,
        conf=0.95, type="all")

# percentile interval:
theta1 <- ci$perc[4]
theta2 <- ci$perc[5]

# basic interval:
theta1 <- ci$basic[4]
theta2 <- ci$basic[5]

# BCa interval:
theta1 <- ci$bca[4]
theta2 <- ci$bca[5]
\end{lstlisting}
%------------------------------------------------------------------------- 

This is somewhat misleading, however, because the difference between the
different confidence intervals is marginal for large $n$ anyway, and
the convergence rate for large $n$ is therefore of merely theoretical
interest. Of practical relevance is instead the behavior for small $n$,
where the bootstrap intervals perform poorer than the classic intervals
(see section \ref{sec:vergleich}). In defense of the comparatively poor
performance of the bootstrap for small $n$ in a specific study, its inventor,
Bradley Efron, wrote \cite{efron87}:
\begin{quote}
  ``Bootstrap methods are intended to supplement rather than replace
  parametric analysis, particularly when parametric methods can't be used
  because of modeling uncertainties or theoretical intractability.''
\end{quote}
The function {\em boot.ci} from the R library {\em boot} can compute a number
of bootstrap confidence intervals, including the three aforementioned.
According to Efron \& Tibshirani \cite{efron86}, the minimum value for the
number $r$ of bootstrap replications should be $R=1000$. Usage of the
R function {\em boot.ci} is shown in listing \ref{lst:bootstrap}.

Apart from the confusion about the most appropriate bootstrap interval in a
particular situation, the bootstrap method has another drawback:
as it is based on Monte Carlo simulations, its results are not deterministic
and not reproducible. This means that two different researchers will end
with different confidence intervals for the same data. Leon Jay Gleser
sees therein a violation of a rule that he calls the ``first law of
applied statistics'' (comments to \cite{diciccio96}, p.~220):
\begin{quote}
``Two individuals using the same statistical
method on the same data should arrive at the
same conclusion.''
\end{quote}
It should be noted that the differences are small, though. Nevertheless
they are noticeable and the bootstrap method might therefore leave some
users with a slightly uneasy feeling.

%-----------------------------------------------------------------------
\begin{figure*}[!t]
  \begin{center}
  \subfigure[Wald interval]{
    \includegraphics[width=0.45\textwidth]{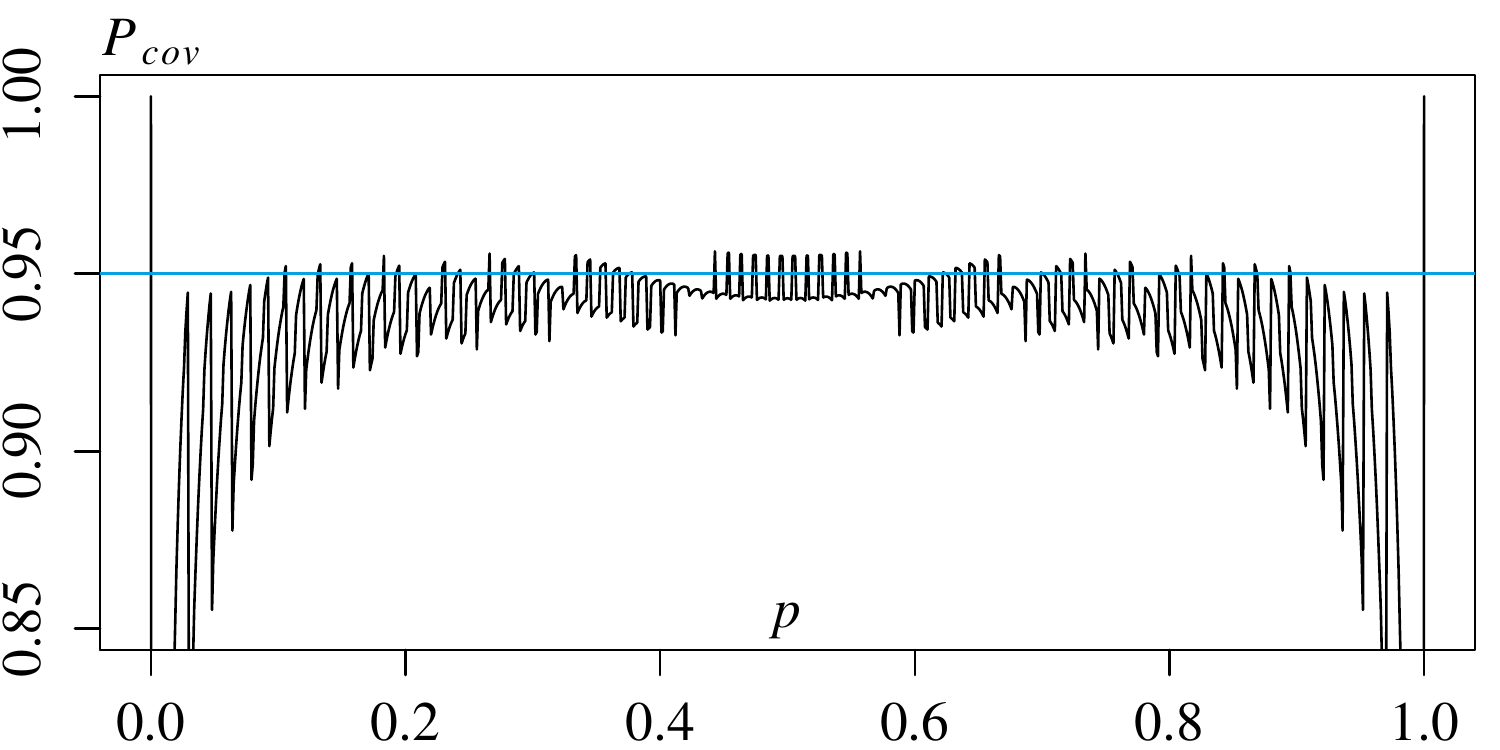}
  }
  \subfigure[Wilson interval]{
    \includegraphics[width=0.45\textwidth]{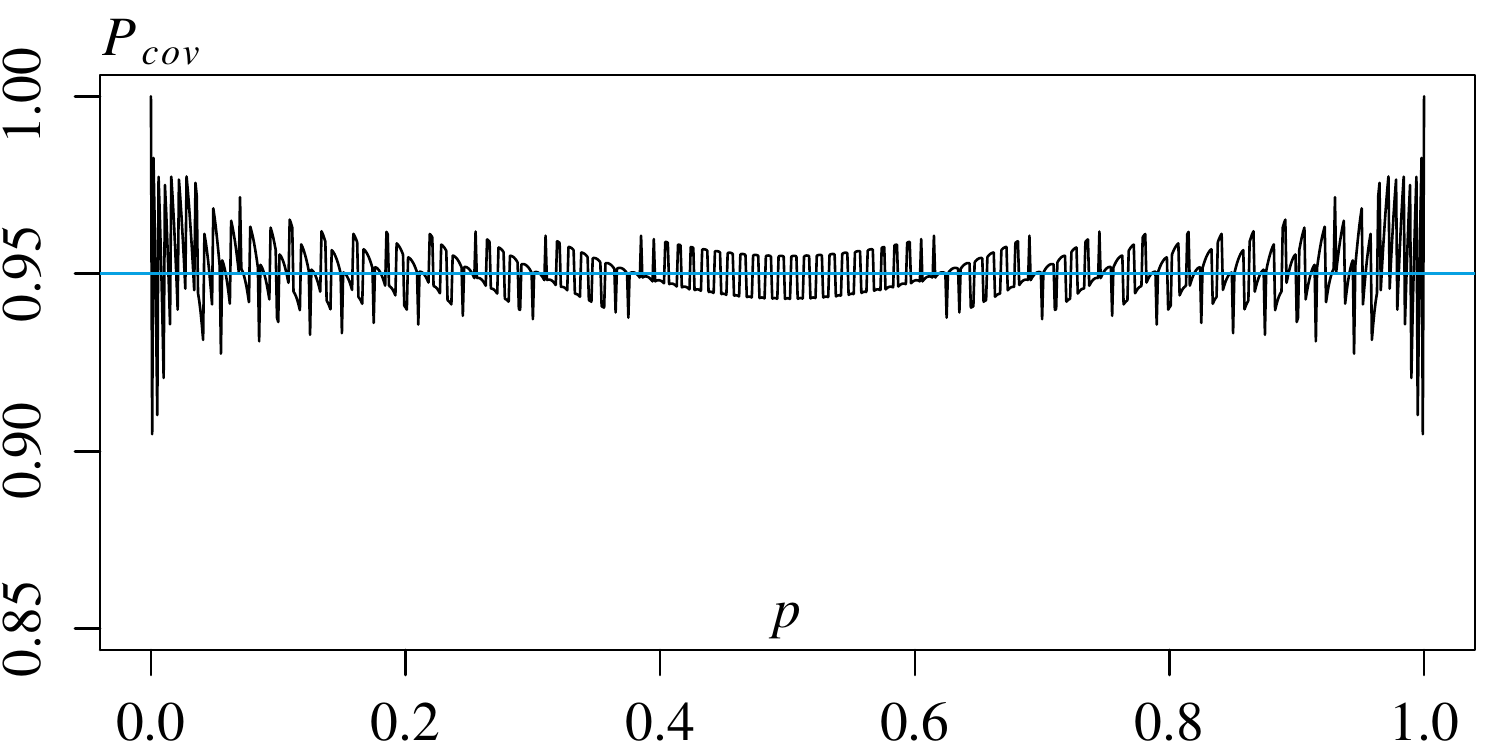}
  }
  \subfigure[LR support interval ($K=8$)]{
    \includegraphics[width=0.45\textwidth]{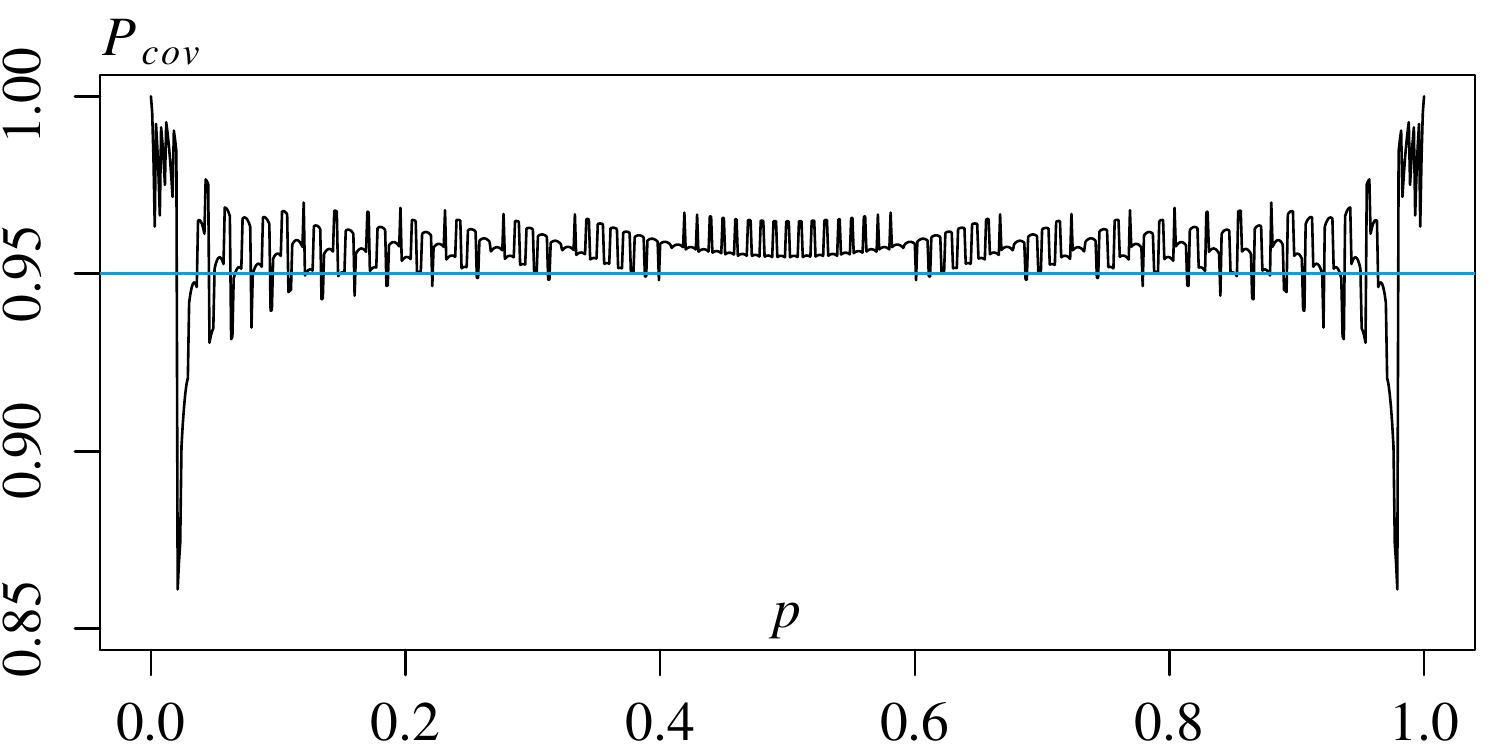}
  }
  \subfigure[HPD interval]{
    \includegraphics[width=0.45\textwidth]{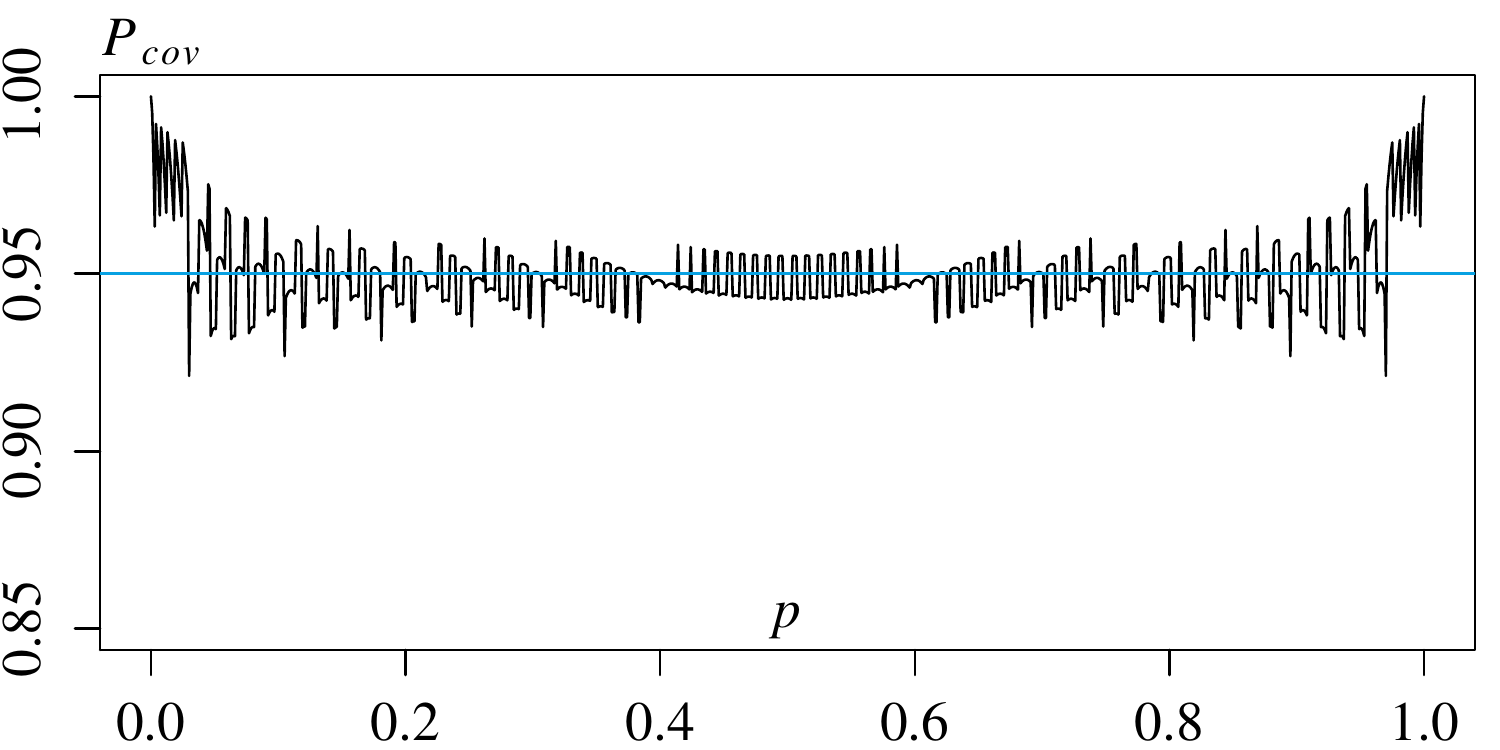}
  }
  \end{center}
\caption{\label{fig:pcov-binom}Coverage probability $P_{cov}(p)$ of the confidence intervals for a binomial proportion as a function of the true parameter value $p$ for $n=100$ and $1-\alpha=0.95$.}
\end{figure*}
%-----------------------------------------------------------------------

\section{Performance in examples}
%-----------------------------------------------------------------------
\label{sec:vergleich}
This section provides a comparative evaluation of the different confidence
intervals with respect to examples for all three of the aforementioned cases.
Apart from the coverage probability $P_{cov}$, the relative size of the
confidence intervals is of interest, too.

For fixed $n$, the relative frequency can only be one of $n+1$ discrete
values, so that $P_{cov}(p)$ can be computed exactly. As an example for
the mean value, I have chosen an asymmetric distribution with density
$f(x)=3x^2$, such that Monte Carlo simulations might show whether the
bootstrap provides any advantages over the classic intervals that assume
symmetry. As an example for an ML estimator, I have chosen the parameter
$\lambda$ of the exponential distribution. In this example, even the inverse
of the Hesse matrix can be calculated analytically in closed form, which allows
for a comparison of all methods by means of a Monte Carlo simulation. From
the bootstrap methods, I have only tested the non-parametric bootstrap because
the parametric method is not universally applicable, but must be tailored to
each particular use case, which might be too much of an effort for an end user only interested in confidence intervals\footnote{Apart from an understanding of probability theory, it also requires knowledge about the generation of random numbers (transformation method, rejection method \cite{press89}).}.

\subsection{$P_{cov}$ for the relative frequency}
%-----------------------------------------------------------------------
\label{sec:vergleich:p}
The coverage probability of different confidence intervals for a binomial
proportion was already investigated by Brown et al.~\cite{brown01}.
Based on their results, they recommended the Wilson interval. As they did not
include the LR support interval or the HPD interval in their study, the
corresponding behavior of $P_{cov}(p)$ as a function of $p$ is shown
in Fig.~\ref{fig:pcov-binom}. The corresponding behavior of the ``exact'' 
(Clopper-Pearson) interval is shown in Fig.~\ref{fig:pcov-binom-exact}. 
The curves have been computed as follows:
\begin{itemize}
\item for every $0\leq k\leq n$, the confidence interval was calculated
\item for every sampled value $p\in[0,1]$, the probabilities of all $k$
  were added for which $p$ fell into the confidence interval
\end{itemize}
As already noted by Brown et al., the classical Wald interval taught in
statistics text books has a way too low coverage probability almost over the
entire range of $p$ values. $P_{cov}$ even approaches zero for small or large
$p$. The Wilson interval, on the contrary, fluctuates around the nominal
value, albeit with greater deviations towards the boundaries of the $p$-range.
Interestingly, the HPD interval has an even better better coverage probability
than the Wilson interval because it mostly shows a similar behavior, but
has no too small values at the boundaries. The behavior of the LR
support interval for $K=8$ is similar to that of the exact Clopper-Pearson
interval, but there are instances where $P_{cov}$ falls considerably below
the nominal value.
%-----------------------------------------------------------------------
\begin{figure}[!b]
\centering\includegraphics[width=0.95\columnwidth]{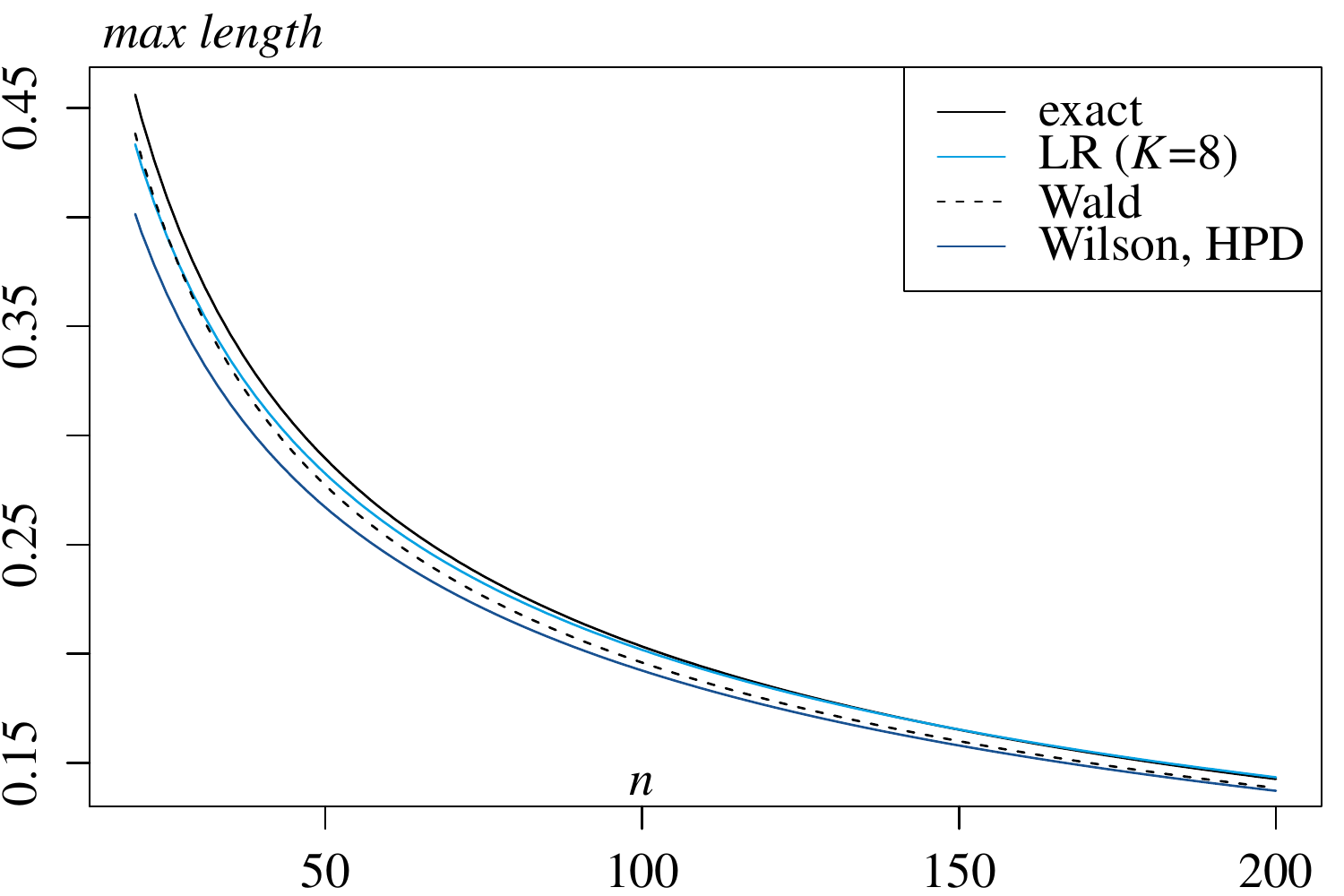}
\caption{\label{fig:maxlength-binom}Maximum length of the confidence intervals for the relative frequency as a function of $n$ for $1-\alpha=0.95$. The maximum length of the HPD and Wilson interval are nearly identical.}
\end{figure}
%-----------------------------------------------------------------------

%-----------------------------------------------------------------------
\begin{figure}[!t]
\centering\includegraphics[width=0.95\columnwidth]{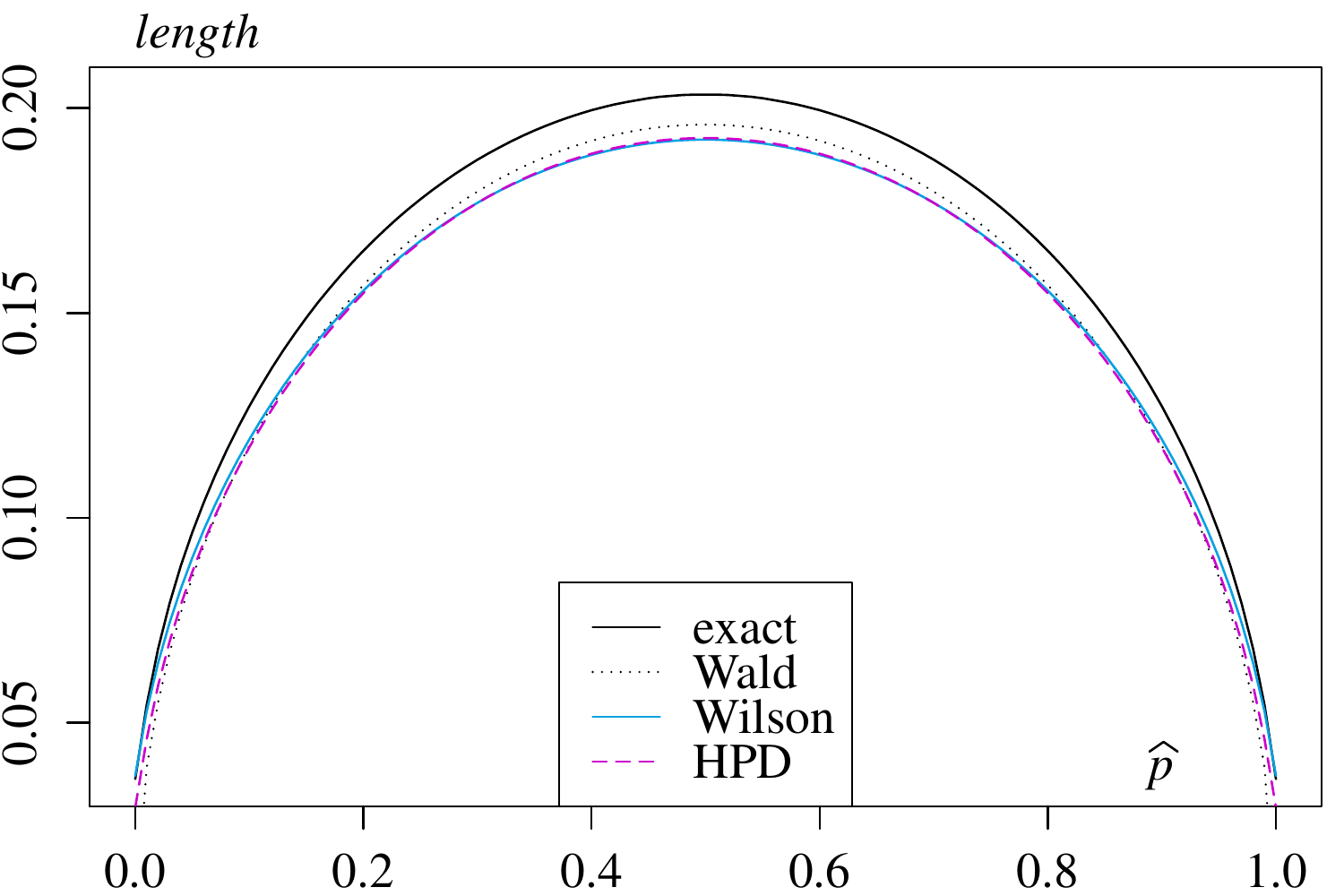}
\caption{\label{fig:length-binom}Confidence interval length for the relative frequency as a function of $\hat{p}$ for $1-\alpha=0.95$ and $n=100$.}
\end{figure}
%-----------------------------------------------------------------------

Another evaluation criterion is the interval length, which should be minimal
for comparable coverage probability. The maximum length of all intervals occurs
for $p=1/2$ and is plotted as a function of $n$ in 
Fig.~\ref{fig:maxlength-binom}. The widest interval is the exact interval,
which is inevitable prize for the guarantee of $P_{cov}(p)\geq 1-\alpha$ with
a greater $P_{cov}$ more often than not. Curiously enough, the maximum length
of the Wald interval is greater than that of the Wilson or HPD interval, 
although its coverage probability is smaller. This apparent contradiction
is resolved when the interval lengths for varying $\hat{p}$ with fixed $n$
are compared (see Fig.~\ref{fig:length-binom}). It can be seen that the 
classical Wald interval is unnecessarily wide for $\hat{p}\approx 0.5$, but
too short for $\hat{p}\approx 0$ or $\hat{p}\approx 1$.

It is interesting to note that the HPD interval for $\hat{p}\approx 0$ or
$\hat{p}\approx 1$ is shorter than the Wilson interval, even though it has
a considerably higher coverage probability in this region. With respect to
the criteria coverage probability and length, the HPD interval has the
best properties. It has the drawback, though, that it can be computed
only numerically (see listing \ref{lst:hpd-binom}). If a closed formula is 
required, the Wilson interval (see Eq.~(\ref{eq:freq:pwilson})) can be 
used as an alternative, provided $\hat{p}$ is not too close to zero or one.

\subsection{$P_{cov}$ for the mean value}
%-----------------------------------------------------------------------
\label{sec:vergleich:mu}
To compare the classical confidence intervals for the mean value with
the bootstrap intervals, I have chosen a random variable with the
probability density
\begin{equation}
  \label{eq:dichtex2}
  f(x) = \left\{ \begin{array}{ll} 3x^2 & \mbox{ for }0\leq x\leq 1 \\
    0 & \mbox{ otherwise}\end{array} \right.
\end{equation}
The expectation value of this distribution is $3/4$, and random numbers
drawn from this distribution can be generated by means of the transformation
method \cite{press89} with
\begin{quote}
  \verb$runif(N, min=0, max=1)**(1/3)$
\end{quote}
The number of simulated mean value measurements was set to $N=10^6$, which
means that the coverage probability can be estimated with an accuracy
$\pm 0.0004$ for $\alpha=0.05$.

%-----------------------------------------------------------------------
\begin{figure}[!t]
  \begin{center}
    \subfigure[Coverage probability]{
      \includegraphics[width=0.95\columnwidth]{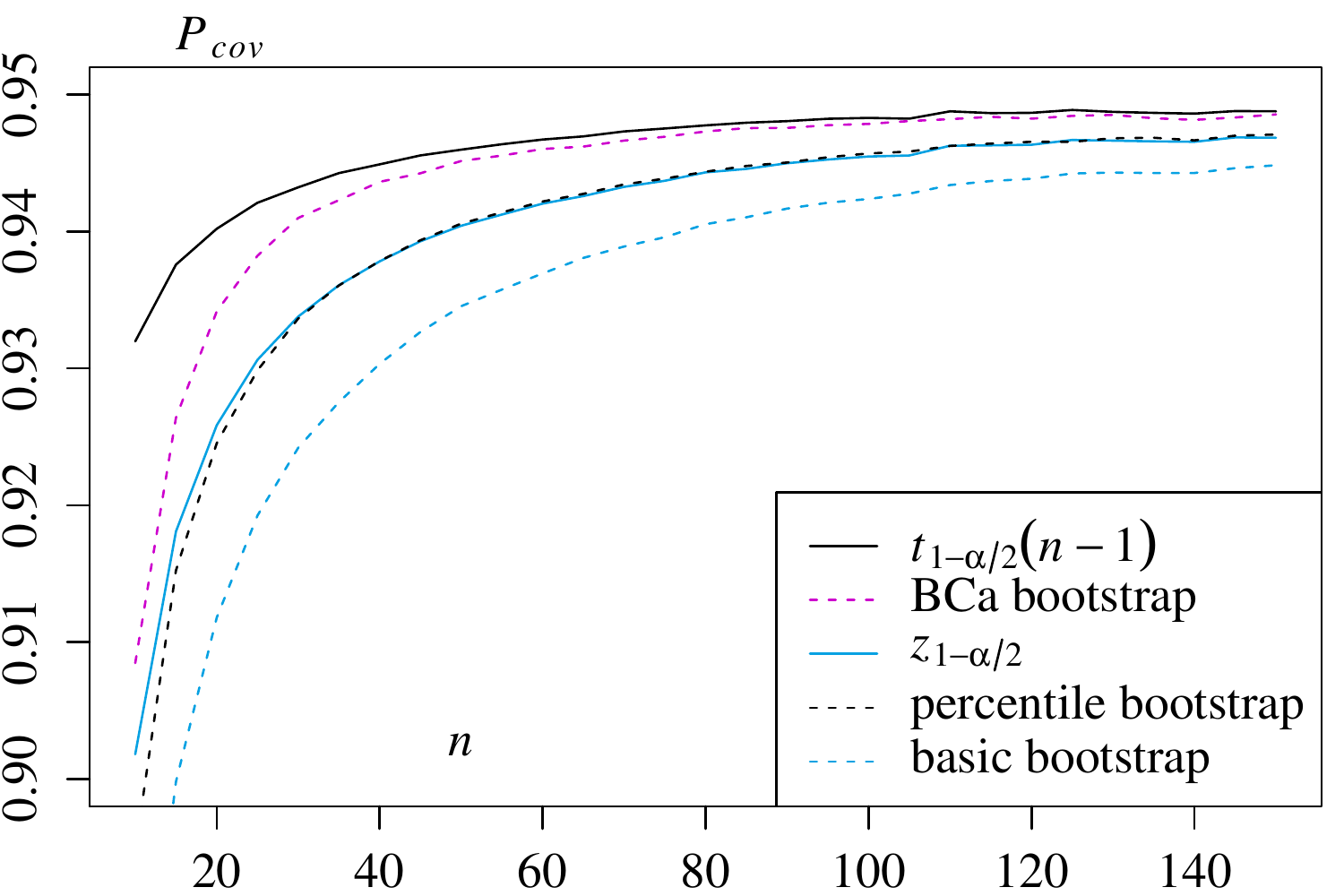}
    }
    \subfigure[Average length]{
      \includegraphics[width=0.95\columnwidth]{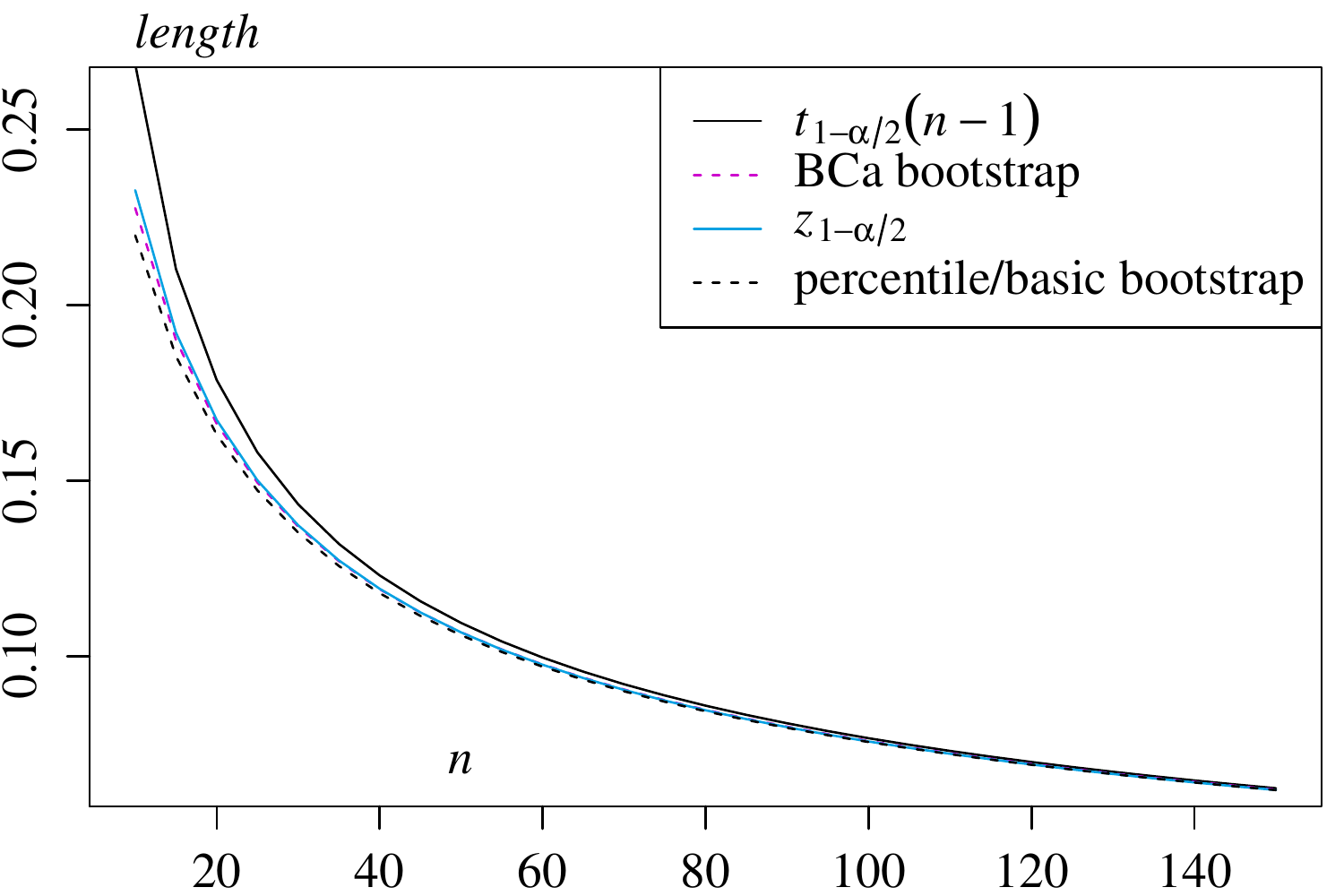}
    }
  \end{center}
\caption{\label{fig:mean-pcov}Coverage probability and average length of the different confidence intervals for the mean value of $n$ random variables distributed according to Eq.~(\ref{eq:dichtex2}).}
\end{figure}
%-----------------------------------------------------------------------

The behavior of $P_{cov}$ and the length of the different confidence intervals
as a function of the number $n$ of observed data points is shown in
Fig.~\ref{fig:mean-pcov}. Surprisingly, the classic confidence interval
based on the $t$ distribution has the best coverage probability throughout,
although the distribution of $\overline{x}$ is asymmetric for small $n$.
The weaknesses of the bootstrap method for small $n$ are thus not compensated
in this case by its ability to take asymmetries into consideration.
The best bootstrap interval in this case is the $BC_a$ interval. It has
a length that is is similar to the classic $z_{1-\alpha/2}$ interval, but
with a greater $P_{cov}$. Venables' \& Ripley's recommendation for
the basic over the percentile bootstrap cannot be confirmed, but, on the
contrary, the basic bootstrap interval has a clearly too low $P_{cov}$
in this case, whilst the percentile interval has a coverage probability that
is comparable to the classic $z_{1-\alpha/2}$ interval.

\subsection{$P_{cov}$ for ML estimators}
%-----------------------------------------------------------------------
\label{sec:vergleich:mu}
Let us consider the exponential distribution as a test case for comparing
the different confidence intervals for maximum likelihood estimators.
The exponential distribution has the single parameter $\lambda$ and 
the probability density
\begin{equation}
  \label{eq:dichteexp}
  f(x) = \left\{ \begin{array}{ll} \lambda e^{-\lambda x} & \mbox{ for }x\geq 0 \\
    0 & \mbox{ otherwise}\end{array} \right.
\end{equation}
The log-likelihood function obtained from this density is
\begin{equation}
  \label{eq:ellexp}
  \ell(\lambda) = n\log(\lambda) + \lambda\sum_{i=1}^n x_i
\end{equation}
The ML estimator for $\lambda$ is obtained by solving the equation
(\ref{eq:ML}) for $\lambda$:
\begin{equation}
  \label{eq:lambda-ML}
  \hat{\lambda} = \frac{n}{\sum_{i=1}^n x_i} = \frac{1}{\overline{x}}
\end{equation}
As the exponential distribution only has a single parameter, the Hessian matrix
is of dimension $1\times 1$, ergo a scalar. It can be readily computed as
\begin{equation}
  \label{eq:exp-HM}
  H(\lambda) = \left(\frac{\partial^2}{\partial\lambda^2}\ell\right)
  = \left(-\frac{n}{\lambda^2}\right)
\end{equation}
When the variance of $\hat{\lambda}$ is estimated form the Hessian with the
method of section \ref{sec:ml:hesse}, it reads
\begin{equation}
  \label{eq:exp-sigma-HM}
  \hat{\sigma}_{\mbox{\scriptsize\em HM}} = \sqrt{\Big(-H(\hat{\lambda})\Big)^{-1}} = \frac{\hat{\lambda}}{\sqrt{n}}
\end{equation}

%-----------------------------------------------------------------------
\begin{figure}[!t]
  \begin{center}
    \subfigure[Coverage probability]{
      \includegraphics[width=0.95\columnwidth]{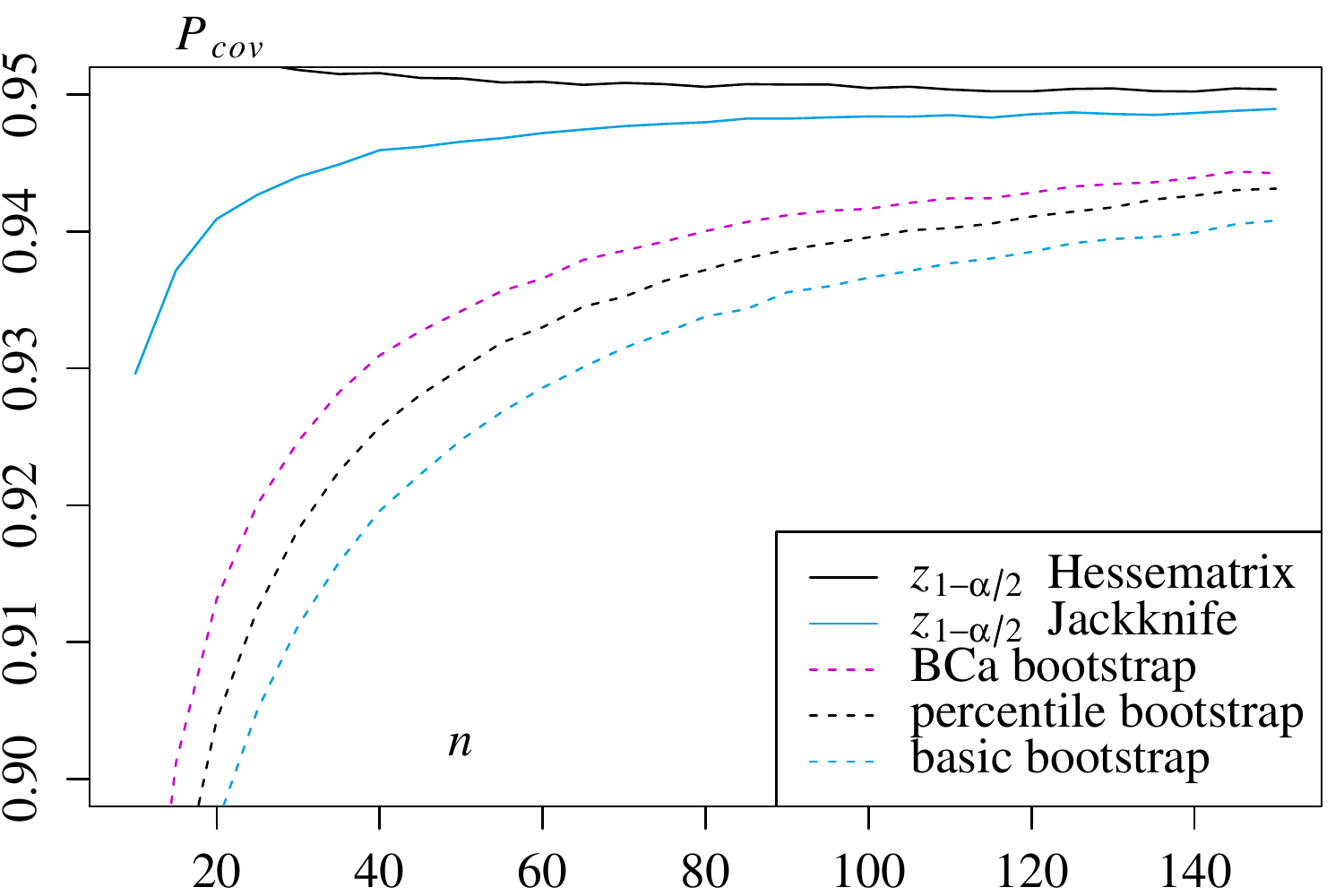}
    }
    \subfigure[Average length]{
      \includegraphics[width=0.95\columnwidth]{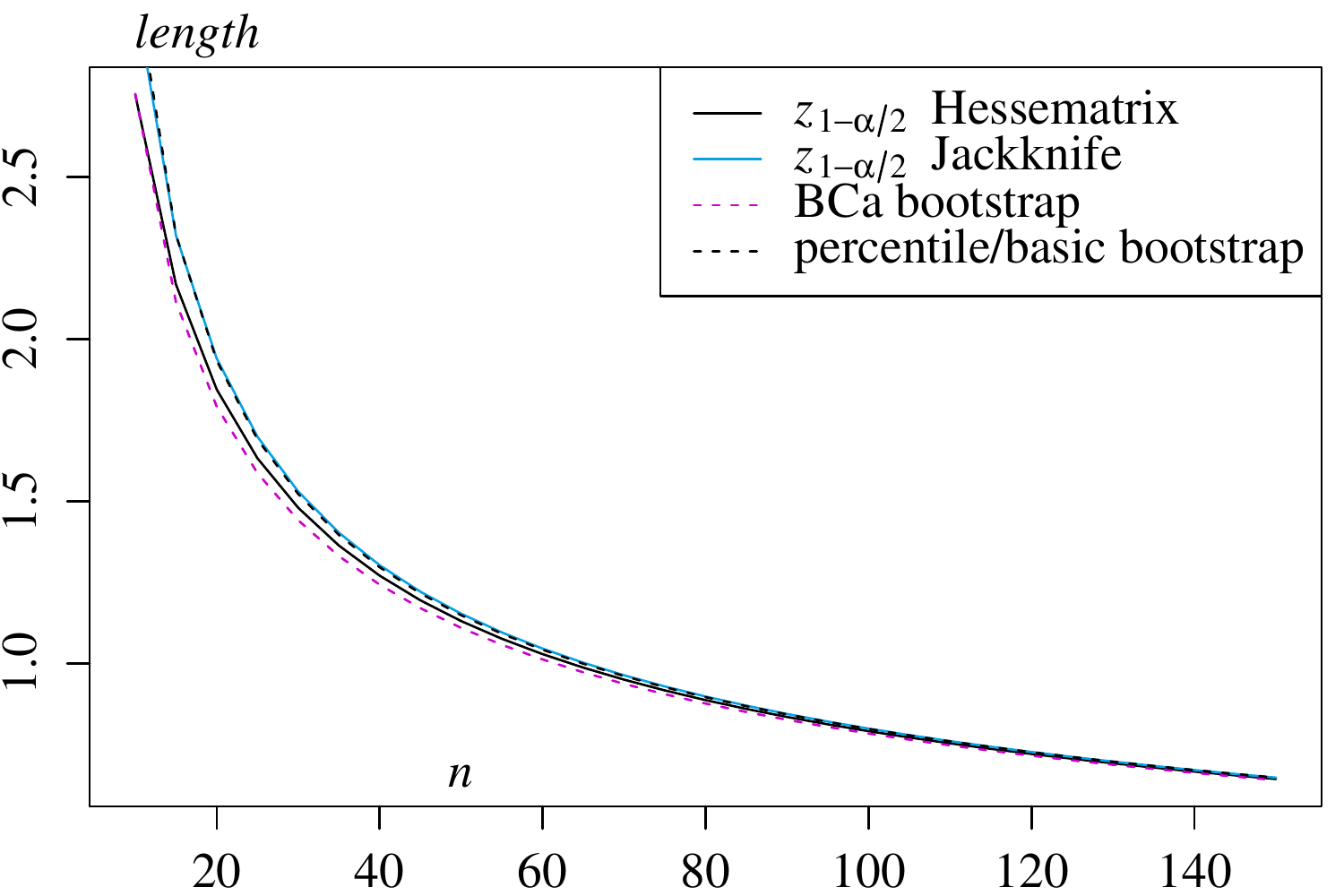}
    }
  \end{center}
\caption{\label{fig:lambda-pcov}Coverage probability and average length of the different confidence intervals for the ML estimator of the parameter $\lambda$ of the exponential distribution.}
\end{figure}
%-----------------------------------------------------------------------

Again, I have generated $N=10^6$ times $n$ exponentially distributed random
numbers with $\lambda=2$ in order to simulate the distribution of
$\hat{\lambda}$ and to compare $P_{cov}$ and average length of the different
confidence intervals. The results are shown in Fig.~\ref{fig:lambda-pcov}.
The classical interval with $\hat{\sigma}_{\mbox{\scriptsize\em HM}}$ has the best
coverage probability, followed by the classical interval with
$\hat{\sigma}_{\mbox{\scriptsize\em JK}}$. Among the bootstrap intervals,
the $BC_a$ interval has the highest coverage probability, and again the
percentile bootstrap performs better than the basic bootstrap.
Venables' \& Ripley's recommendation in favor of the basic bootstrap must
therefore be rejected. Overall, the bootstrap intervals show a 
coverage probability that is clearly below the nominal value $1-\alpha$.

It is surprising that the confidence interval based on the jackknife variance
is wider, but has a smaller coverage probability than the interval based
on the Hessian. A closer look at the simulated distribution of 
$\hat{\lambda}$ reveals that in this case the ML estimator is biased and is on average too large\footnote{ML estimators are only guaranteed to be {\em asymptotically} unbiased for large $n$.}.
As $\hat{\sigma}_{\mbox{\scriptsize\em HM}}$ is proportional to $\hat{\lambda}$
according to Eq.~(\ref{eq:exp-sigma-HM}), the confidence interval is wider
when the estimated value is too large, which compensates the bias of the ML
estimator in this case. This lead to a correlation of 
$|\hat{\lambda}-\lambda|$ with respect to  
$\hat{\sigma}_{\mbox{\scriptsize\em HM}}$ of about 
$0.60$ in the Monte-Carlo simulations for $n=20$, but of only about $0.40$
with respect to $\hat{\sigma}_{\mbox{\scriptsize\em JK}}$. This explains why 
$P_{cov}$ can be smaller for the wider interval.

\section{Conclusions}
%-----------------------------------------------------------------------
\label{sec:fazit}
For the practitioner, the comparative evaluation of the different confidence
intervals leads to the following recommendations:
\begin{enumerate}
\item For a relative frequency, the HPD interval (listing \ref{lst:hpd-binom})
  or the Wilson interval (Eq.~(\ref{eq:freq:pwilson})) should be used.
  The Wilson interval has the advantage of a closed formula, but it has a
  smaller coverage probability than the HPD interval for $p$ values
  close to zero or one.
\item For mean values, the classical confidence interval based on the $t$ 
  distribution should be used (Eq.~(\ref{eq:freq:mu-t})).
\item For ML estimators with a smooth log-likelihood function,
  the confidence interval $z_{1-\alpha/2}\cdot\hat{\sigma}$ should be used.
  The variance $\hat{\sigma}$ can be estimated either from the Hessian matrix
  or, in a simpler way, by means of the jackknife (listing \ref{lst:jackknife}).
\item In the remaining cases, the $BC_a$ bootstrap interval should be used.
\end{enumerate}
The results of this technical report thus confirm the already cited remark
by Efron \cite{efron87}:
\begin{quote}
  ``Bootstrap methods are intended to supplement rather than replace
  parametric analysis, particularly when parametric methods can't be used
  because of modeling uncertainties or theoretical intractability.''
\end{quote}

\bibliographystyle{ieeetr}
\bibliography{confidence-intervals-en}

\end{document}